\documentclass[11pt,fleqn]{article}       
\usepackage{geometry}
\usepackage[T1]{fontenc}
\usepackage[utf8]{inputenc}
\usepackage{authblk}
\usepackage{mathrsfs}
\usepackage[font=small,labelfont=bf]{caption}
\usepackage{graphicx}
\usepackage{multirow}
\usepackage{amsmath,amssymb,amsfonts}
\usepackage{verbatim}
\usepackage{bm}
\usepackage{color}
\usepackage{ulem}
\usepackage{mathtools}
\usepackage{cite}
\usepackage{colortbl}
\usepackage{cancel}
\definecolor{light-gray}{gray}{0.85}
\usepackage{hyperref}
\hypersetup{
   bookmarksnumbered,
   citecolor={blue},
   linkcolor={blue},
   urlcolor={blue},
   filecolor={blue}
} 

\newcommand{\be}{\begin{equation}}
\newcommand{\ee}{\end{equation}}
\newcommand{\ba}{\begin{eqnarray}}
\newcommand{\ea}{\end{eqnarray}}
\def\la{\mathrel{\mathpalette\fun <}}
\def\ga{\mathrel{\mathpalette\fun >}}
\def\fun#1#2{\lower3.6pt\vbox{\baselineskip0pt\lineskip.9pt
        \ialign{$\mathsurround=0pt#1\hfill##\hfil$\crcr#2\crcr\sim\crcr}}}

\def\blue{\textcolor{blue}}

\newcommand{\AddIPAC}{%
\textit{IPAC, California Institute of Technology, Mail Code 314-6, 1200 E. California Blvd., Pasadena, CA 91125}
}

\newcommand{\AddrTexas}{%
\textit{Department of Physics, The University of Texas at Austin, Austin, 78712 TX, USA}
}
\newcommand{\AddrStockholm}{
\textit{Oskar Klein Center for Cosmoparticle Physics, University of Stockholm, 10691 Stockholm, Sweden}
}
\newcommand{\AddrNordita}{
\textit{Nordita, KTH Royal Institute of Technology and Stockholm University, 10691 Stockholm, Sweden}
}

\usepackage{fancyhdr}
 \usepackage[bottom]{footmisc}
\fancypagestyle{plain}{%
    \fancyhead[R]{UT-WI-13-2025, NORDITA 2025-TBD}
    
}
\date{}
\title{\Large\bf Model-Independent Dark Energy Measurements from DESI DR2 and Planck 2015 Data}
\author[1]{Yun Wang\thanks{wang@ipac.caltech.edu}}
\author[2,3,4]{Katherine Freese\thanks{ktfreese@utexas.edu}}
\affil[1]{\AddIPAC}
\affil[2]{\AddrTexas}
\affil[3]{\AddrStockholm}
\affil[4]{\AddrNordita}

\begin{document}
\maketitle
\vspace*{0mm}
\begin{abstract}
Using DESI DR2 baryon acoustic oscillation (BAO) distance measurements and Planck cosmic microwave background distance priors, we have measured the dark energy density $\rho_X(z)$ and dark energy equation of state $w_X(z)$ as free functions of redshift (smoothly interpolated from values at $\{z_i\}=\{0, 1/3, 2/3, 1, 4/3, 2.33\}$), and find both to be consistent with a cosmological constant, with only deviations of $\sim 1\sigma$ for $\rho_X(z)$ and $\sim 2 \sigma$ for $w_X(z)$ at $z=2/3$. We also find that measuring $\{\rho_X(z_i)\}$ is preferred to measuring $\{w_X(z_i)\}$ by model selection using the Akaike Information Criterion (AIC) as well as the Bayesian Information Criterion (BIC); we confirm our earlier finding  in Wang \& Freese (2006) that $w_X(z)$ is significantly less constrained by data than $\rho_X(z)$. 
We show that varying the choice of redshift values of the $\rho_X(z)$ measurements leads to very consistent results, with AIC/BIC slightly favoring the case of our fiducial redshifts $\{z_i\}$ but with $z=4/3$ omitted. 
We find agreement with a cosmological constant except for the 1-2$\sigma$ deviation at $0.4 \la z \la 0.9$, where DESI DR2 BAO measurements deviate from a cosmological constant at similar statistical significance. 

Our results differ noticeably from those of the DESI Collaboration, in which they used the same DESI DR2 data combined with Planck data and found a 3.1$\sigma$ deviation from a cosmological constant, a finding which is primarily the consequence of their assuming the parametrization $w_X(z)=w_0+w_a(1-a)$.
Our results indicate that assuming a linear $w_X(z)$ could be misleading and precludes discovering how dark energy actually varies with time at higher redshifts.
In our quest to discover the physical nature of dark energy, the most urgent goal at present is to determine definitively whether dark energy density varies with time. We have demonstrated that it is of critical importance to measure dark energy density as a free function of redshift from data. 
Future galaxy redshift surveys by Euclid and Roman at higher redshifts will significantly advance our understanding of dark energy.
\end{abstract}
\clearpage

\section{Introduction}

Twenty-seven years after the discovery of cosmic acceleration 
\cite{Riess98,Perl99}, 
its physical cause remains a mystery. We still cannot differentiate between an unknown energy density and the modification of general relativity as the explanation for cosmic acceleration, but its unknown cause is generally referred to as ``dark energy" (see \cite{Wang-DE-book} for a textbook introduction). Given our ignorance of the nature of dark energy, it is critically important to derive model-independent constraints on dark energy from observational data, to ensure that we are asking the right questions about dark energy.

Galaxy clustering (using data from galaxy redshift surveys) provides a robust probe of cosmic acceleration by enabling the distance measurements from baryon acoustic oscillations (BAO) \cite{Blake2003,Seo2003} sensitive to cosmic expansion history, and large-scale structure growth rate measurements from linear redshift space distortions (RSD) \cite{Kaiser1987} which enable the test of general relativity \cite{Guzzo2008,Wang2008a}.
Recently, the DESI Collaboration found that DESI DR2 BAO data combined with Planck Cosmic Microwave Background (CMB) data lead to 3.1$\sigma$ deviation from a cosmological constant \cite{DESI-DR2}, under two assumptions: (i)  a flat Universe, and (ii)  the dark energy equation of state can be described by the popular parametrization $w_X(z)=w_0+w_a(1-a)$ (with $a$ denoting the cosmic scale factor), the Taylor expansion of $w_X(z)$ keeping the leading term only.
In a second paper \cite{DESI:2025fii}, the DESI collaboration extended their analysis to a variety of alternative approaches and claimed that these other approaches all validate conclusions in \cite{DESI-DR2}. 
However, in this paper, we will show that we find a lower statistical significance for a deviation from a cosmological constant, using the same DESI DR2 + CMB data sets as the DESI collaboration but without assuming the linear parametrization of the dark energy equation of state. 

Instead, we directly obtain the dark energy density from the data as a free function as described below.
As previously shown by Wang and Freese (2006) \cite{WangFreese2006} and reviewed below, measuring the dark energy density function is more direct
and accurate than taking the intermediate step of extracting the dark energy equation of state from data. 
We will comment on comparison of our work with the DESI extended analysis paper \cite{DESI:2025fii} in the discussion section below. 
Our approach is extremely simple and nonparametric and finds lower statistical significance for time variation of dark energy than was found by the DESI Collaboration.

In this paper, we measure the dark energy density $\rho_X(z)$ and $w_X(z)$ as free functions of redshift $z$, parametrized by their values at $\{z_i\}=\{0, 1/3, 2/3, 1, 1.33, 2.33\}$, spanning the redshift range of DESI BAO distance measurements. We also investigate the impact of varying $\{z_i\}$ on the measurment of $\rho_X(z)$.
We present methodology in Sec.\ref{sec:method}, and analysis results in Sec.\ref{sec:results}. We compare our work with that of the DESI Collaboration in Sec.\ref{sec:compare}, and conclude in Sec.\ref{sec:summary}.

\section{Methodology}
\label{sec:method}

\subsection{Basic equations}

Since our main goal is to compare with the DESI DR2 findings \cite{DESI-DR2} in a flat Universe, we assume a flat Universe as well. 
For simplicity, we also fix the neutrino parameters following the usual practice as they are unconstrained by DESI DR2 data. We assume the effective number of neutrino species to be $N_{\rm eff}=3.04$, and the sum of neutrino masses to be 0.06$\,$eV (converted to $\Omega_\nu=\sum_i m_\nu^i/93.14\,{\rm eV}$). The basic equations are:
\ba
\label{eq:basics}
&& D_M(z)\equiv  r(z) = c H_0^{-1} \int_0^z {\rm d}z'\,\frac{1}{E(z')} \\\nonumber
&& D_H(z) \equiv c/H(z)=cH_0^{-1}/E(z) \\\nonumber
&& D_V(z) \equiv \left[z \,D_M^2(z)\, D_H(z)\right]^{1/3}
= cH_0^{-1} \left\{ \frac{z}{E(z)}\left[\frac{r(z)}{cH_0^{-1}}\right]^2\right\}^{1/3}\\\nonumber
&& 
E^2(z) = \Omega_\mathrm{bc}(1+z)^3+\Omega_{\gamma} (1+z)^4
+\Omega_\nu\, \frac{\rho_\nu(z)}{\rho_{\nu, 0}}
+\Omega_{\mathrm X}X(z)
\ea
where $r(z)$ is the comoving distance; $H(z)$ is the Hubble parameter (cosmic expansion rate), $H_0$ is the Hubble constant today,
$ E(z)\equiv H(z)/H_0$; and $D_V(z)$ is related to the spherically averaged distance. 
Note that energy density fraction $\Omega_{\mathrm{bc}}=\Omega_\mathrm{b}+\Omega_\mathrm{c}$, with the subscript "b" and "c" denoting baryons and cold dark matter.
$\Omega_\gamma$, $\Omega_\nu$, and $\Omega_\mathrm{X}$
denote the energy density fractions of radiation, neutrinos, and dark energy, respectively.
Since we are assuming a flat Universe, 
$\Omega_\mathrm{X}=1-\Omega_\mathrm{bc}-\Omega_\gamma-\Omega_\nu$.
Neutrinos transition from radiation at very high $z$ to matter at lower $z$, with the transition occurring at $(1+z) \sim m_\nu/(5\times 10^{-4}\mathrm{eV})$ \cite{ParticleDataGrp2022,Neutrino-review2012},
i.e., $z\sim 120 \, \left(m_\nu/0.06\,\mathrm{eV}\right)$.
We define
\begin{equation}
X(z)\equiv \rho_X(z)/\rho_X(0) \, 
\end{equation} 
the ratio of the dark energy density at redshift $z$ to its value today,
where subscript $X$ refers to the dark energy component of the Universe.

For comparison with DESI DR2 data \cite{DESI-DR2}, we compute $r_d$, the sound horizon at the baryon drag epoch (i.e., the BAO scale), the same way as \cite{DESI-DR2}, by using Eq.(2) in \cite{DESI-DR2}:
\be
r_d = 147.05\,\mbox{Mpc} 
\left(\frac{\omega_\mathrm{b}}{0.02236}\right)^{-0.13}
\left(\frac{\omega_\mathrm{bc}}{0.1432}\right)^{-0.23}
\left(\frac{N_{\mathrm{eff}}}{3.04}\right)^{-0.1},
\ee
where $\omega_{\mathrm{b}}\equiv \Omega_{\mathrm{b}} h^2$,
and $\omega_{\mathrm{bc}} \equiv \Omega_{\mathrm{bc}}h^2$. 
The dimensionless Hubble constant $h$ is defined by $h \equiv H_0/(100\, {\rm km}\,\,{\rm s}^{-1}\,{\rm Mpc}^{-1})$.
$N_{\mathrm{eff}}$ is the effective number of neutrino species. Following \cite{DESI-DR2}, we assume the standard neutrino content of $N_{\mbox{eff}}=3.04$, and the sum of neutrino masses $\sum m_\nu = 0.06\,$eV, which are also the assumptions made in \cite{WangDai2016} in deriving the CMB distance priors.

We measure both $X(z)$ and $w_X(z)$ from data. They are related by
\be
X(z) \equiv \frac{\rho_X(z)}{\rho_X(0)} = \exp\left[ \int_0^z {\rm d}z'\,\,\frac{1+w_X(z')}{1+z'}\right]
\label{eq:wX}
\ee

The set of parameters we estimate are the cosmological parameters $\{\Omega_m h^2, h, \Omega_b h^2\}$ and dark energy parameters, which are $X(z_i)$ ($i=2,3,...,6)$
for $\rho_X(z)$ (note that $X(0)=1$), and $w_X(z_i)$ ($i=1,2,3,...,6$) for $w_X(z)$:
\ba
&& z_i =(i-1)/3,  \hspace{0.1in} i=1, 2, 3, 4, 5;  \hspace{0.1in} z_6=z_{\rm max}=2.33 \\
&&X(z)= \left\{
\begin{array}{l}
1, \hspace{0.1in}z=z_1=0 \\
\mbox{interpolated via cubic-spline from}\,\,
\{X(z_i)\} (i=1,2,...,6), 
\hspace{0.1in} 0< z < z_{\rm max}\\
X(z_{\rm max}), \hspace{0.1in}z \geq z_{\rm max}
\end{array}
\right.
\label{eq:Xz}\\
&&w(z)= \left\{
\begin{array}{l}
\mbox{interpolated via cubic-spline from} \,\,\{w_X(z_i)\} (i=1,2,...,6), 
\hspace{0.1in} 0\leq z < z_{\rm max}\\
w_X(z_{\rm max}), \hspace{0.1in}z \geq z_{\rm max}
\end{array}
\right.
\label{eq:wz}
\ea
We use cubic spline for interpolation, the standard technique for obtaining a smooth function from a set of discrete measurements.
In addition, we impose $X(z>2.33)=X(2.33)$ and $w_X(z>2.33)=w_X(2.33)$ for convergence where there is no data. 
Fixing $X(z>z_{\rm max})$ and $w_X(z> z_{\rm max})$ reflects the limit of current data (with $z_{\rm max}=2.33)$, and
avoids making assumptions about dark energy at higher redshifts that
can be propagated into artificial constraints on dark energy
at lower $z$ \cite{WangTegmark2004,WangPia2007}.

For reference, the popular linear parametrization for $w_X(z)$ and its corresponding $X(z)$ are:
\ba
&& w_X(z)=w_0+w_a\left(\frac{z}{1+z}\right)  = w_0 + w_a (1-a) \\
&& X(z) = (1+z)^{3(1+w_0+w_a)}\,\exp\left( -\frac{3w_a z}{1+z}\right) 
\label{eq:Xz_w0wa}
\ea
where $a$ is the cosmic scale factor, $a =1/(1+z)$.

\subsection{DESI data used}

We use the DESI DR2 BAO distance measurements from the first seven rows (excluding the labels row) in Table IV of \cite{DESI-DR2}: \\
\noindent
(1) $D_\mathrm{V}/r_\mathrm{d}$ at $z=0.295$ (BGS), 1st row, 5th column.\\
(2) $D_\mathrm{M}/r_\mathrm{d}$ (8th column) and $D_\mathrm{H}/r_\mathrm{d}$ (9th column), and their correlation (10th column), at $z=0.51$ (LRG1), 0.706 (LRG2), 0.934 (LRG3+ELG1), 1.321 (ELG2), 1.484 (QSO), 2.33 (Ly$\alpha$); rows 2-7.

The correlation of $D_\mathrm{M}/r_\mathrm{d}$ and $D_\mathrm{H}/r_\mathrm{d}$ are included in the analysis by constructing the covariance matrix of $D_\mathrm{M}/r_d$ and $D_\mathrm{H}/r_\mathrm{d}$, with the off-diagonal element given by $r_\mathrm{M,H}$.

\subsection{Planck CMB distance priors}

DESI DR2 data alone does not place meaningful constraints on dark energy. We add either CMB distance priors or wide priors on $\{\Omega_m h^2, h, \Omega_b h^2\}$,
where $\Omega_\mathrm{m}$ is the total matter density fraction, following \cite{WangPia2007,WangDai2016}.
The CMB distance priors were first introduced by one of us\cite{WangPia2007} as an effective way to compress cosmological information from CMB data, and are independent of the dark energy model assumed. When combined with other data, the CMB distance priors and the full set of CMB data give very similar results on dark energy parameters (for further discussion of this point, see Section 3.3 and Table 3). 
Note that the full CMB data by itself only place weak constraints on dark energy parameters, which are also sensitive to the prior range of these parameters; thus one must be cautious in using the full CMB data in constraining dark energy, to avoid biasing estimates on dark energy parameters. In contrast, the CMB distance priors are robust, and do not bias the dark energy estimates.

The CMB shift parameters are defined as \cite{WangPia2007}
\ba
\label{eq:R-la}
&& R = \sqrt{\Omega_m H_0^2}\,\, r(z_*)/c \\ \nonumber
&& l_a = \pi r(z_*)/r_s(z_*)
\ea
where $z_*$ is the redshift to the photon-decoupling surface.
The comoving sound horizon at redshift $z$ is given by
\ba
\label{eq:rs}
r_s(z)  &= & \int_0^{t} \frac{c_s\, dt'}{a}
=cH_0^{-1}\int_{z}^{\infty} dz'\,
\frac{c_s}{E(z')}, \nonumber\\
 &= & cH_0^{-1} \int_0^{a} 
\frac{da'}{\sqrt{ 3(1+ \overline{R_b}\,a')\, {a'}^4 E^2(z')}}.
\ea
For $z\geq z_*$, $a^4 E^2(z)=\Omega_\mathrm{bc} a+ \Omega_{\rm rad}+(1-\Omega_m) X(z) a^4$,
$\Omega_{\rm rad}=\Omega_\mathrm{bc} a_{\rm eq}$, and
$z_{\rm eq}=2.5\times 10^4 \Omega_\mathrm{bc} h^2 (T_{CMB}/2.7\,{\rm K})^{-4}$, since neutrinos become non-relativistic at $z\sim 120\, (m_\nu/0.06\,\mathrm{eV})$ \cite{Neutrino-review2012}.
The sound speed is $c_s=1/\sqrt{3(1+\overline{R_b}\,a)}$,
with $\overline{R_b}\,a=3\rho_b/(4\rho_\gamma)$,
$\overline{R_b}=31500\Omega_bh^2(T_{CMB}/2.7\,{\rm K})^{-4}$.
We take $T_{CMB}=2.7255$K.

\subsubsection{Planck 2015 distance priors}

We primarily use the Planck 2015 distance priors on $\{R,l_a,\Omega_bh^2,n_s\}$ for a flat Universe from Eq.(15) and Eq.(16) in \cite{WangDai2016}, marginalizing over $n_s$ (the scalar spectral index, which is the power law index of the primordial matter power spectrum) since we are only using DESI DR2 BAO distance measurements which do not depend on $n_s$.

The redshift to the photon-decoupling surface, $z_*$, is given by the 
fitting formula \cite{Hu96}:
\ba
\label{eq:z*}
&&z_*=1048\, \left[1+ 0.00124 (\Omega_b h^2)^{-0.738}\right]\,
\left[1+g_1 (\Omega_m h^2)^{g_2} \right],\\\nonumber
&&g_1 = \frac{0.0783\, (\Omega_b h^2)^{-0.238}}
{1+39.5\, (\Omega_b h^2)^{0.763}}\\\nonumber
&&g_2 = \frac{0.560}{1+21.1\, (\Omega_b h^2)^{1.81}}
\ea

The Planck CMB priors $\{R,l_a,\Omega_bh^2\}$ are included as correlated Gaussian priors, see Eq.(18) in \cite{WangDai2016}. In order to investigate whether the Planck CMB distance priors bias the dark energy measurements, we also consider two additional cases in which we replace the Planck CMB distance priors with a wide flat prior on $\Omega_m h^2$ for comparison. 

It is worth noting that the CMB distance priors compress CMB data by deriving constraints on $\{R,l_a,\Omega_bh^2\}$ from Planck MCMC chains using a fixed set of formulae, Eqs.(\ref{eq:R-la}-(\ref{eq:z*}) \cite{WangPia2007,WangDai2016}, with $m_\nu=0$, which can be easily computed for any given model.

\subsubsection{Planck 2018 distance priors allowing neutrino mass to vary}

In order to check the robustness of our results, we also use the Planck 2015 and 2018 distance priors on $\left\{R, l_a, \Omega_b h^2, \Omega_ch^2, N_\mathrm{eff} \right\}$ from \cite{Zhai2020}, which allow neutrino mass and $N_\mathrm{eff}$ to vary.
We use Eq.(4.1) and Eq.(4.2) from \cite{Zhai2020}, valid for a flat Universe with massive neutrinos. We marginalize over $N_\mathrm{eff}$, as it is not a parameter in our analysis, but we allow it to vary to test the robustness of our results on $X(z)$. This should lead to less stringent results than using the Planck 2015 priors from \cite{WangDai2016}, which were derived from Planck MCMC chains with $m_\nu=0.06\,$eV.

Since the Planck 2015 and 2018 distance priors (allowing $m_\nu$ to vary) from \cite{Zhai2020} are derived using CAMB to compute $z_*$, we use the fitting formula for $z_*$ from \cite{Aizpuru2021}, derived using generic algorithms, and accurate to 0.0005\% \cite{Aizpuru2021}.
For fixed $\Omega_\mathrm{m}$, $r(z_*)$ differs by $\sim$ 0.025\% between $m_\nu=0.06\,$eV and $m_\nu=0$, thus $r(z_*)$ can be computed using Eq.(\ref{eq:basics}) with $\Omega_\nu$ absorbed into $\Omega_\mathrm{m}$.
The sound horizon at $z_*$, $r_s(z_*)$, can be computed using Eq.(\ref{eq:rs}).

The Planck 2018 distance priors from \cite{Zhai2020} have not been broadly used, perhaps because \cite{Zhai2020} specifies the requirement of running a full CMB code such as CAMB to compute $r(z_*)$. 
With our simple prescription above, the Planck 2015 and 2018 distance priors (allowing $m_\nu$ to vary) from \cite{Zhai2020} can be conveniently computed with sufficient accuracy, without running a full CMB code.

Note that allowing $m_\nu$ and $N_\mathrm{eff}$ to vary, Planck 2015 distance priors prefer $N_\mathrm{eff}=3.0676$,
while Planck 2018 distance priors prefer $N_\mathrm{eff}=2.8979$
\cite{Zhai2020}, which may indicate that the Planck 2015 distance priors are more reliable since the standard value is $N_\mathrm{eff}=3.04$. 

\subsection{Model selection}

To evaluate the various cases we have studied in our primary analysis for measuring $\rho_X(z)$ and $w_X(z)$ as free functions,  we use the Akaike Information Criterion (AIC) \cite{Akaike1974} and the Bayesian Information Criterion (BIC) \cite{Schwarz1978} for model selection:
\ba
&& \mbox{AIC} \equiv -2 \,\mbox{ln}(L) + 2 N_{\mbox{par}} \\\nonumber
&& \mbox{BIC} \equiv -2 \,\mbox{ln}(L) + N_{\mbox{par}} \mbox{ln}\left(N_{\mbox{data}}\right),
\ea
where $L=\exp(-\chi^2/2)$ is the maximized likelihood of the model,
$N_{\mbox{par}}$ and $N_{\mbox{data}}$ are the numbers of model parameters and data points respectively.
The models with the lowest AIC and BIC values are generally preferred.  When comparing two models, the difference in the values of these quantities between the two, $\Delta\mbox{AIC}$ or $\Delta\mbox{BIC}$, determines 
the strength of the evidence  favoring the model with the lower AIC or BIC over the other, as follows:
$0< \Delta\mbox{AIC} < 2$ or $0 < \Delta \mbox{BIC} < 2$ indicate weak evidence; $2 \leq \Delta\mbox{AIC} < 4$ or $2 \leq \Delta \mbox{BIC} < 6$ indicate moderate evidence; and 
larger $\Delta\mbox{BIC}$ and $\Delta \mbox{BIC}$ values indicate strong evidence, with $\Delta\mbox{AIC}>10$ or $\Delta \mbox{BIC} > 10$ representing very strong evidence favoring the model with the lower AIC or BIC values.

\section{Analysis Results}
\label{sec:results}

The set of parameters in our MCMC analysis is $\{\Omega_m h^2, \Omega_b h^2, h, X(z_{i+1}) \}$ ($i=1,2,...,N$) for measuring the dark energy density $\rho_X(z)$ (note that $X(z)\equiv\rho_X(z)/\rho_X(0)$).  Further, the set of parameters in our MCMC analysis for measuring the dark energy density equation of state $w_X(z)$ is
$\{\Omega_m h^2, \Omega_b h^2, h, w_X(z_{i}) \}$ ($i=1,2,...,N$).  The study of $w_X(z)$ requires one additional parameter compared to the study of $X(z)$ at the same redshifts, the value of $w_X(z=0)$.
On the other hand,  $X(z=0) = 1$ (i.e., its value at z=0 is fixed). 
In our MCMC analysis, we have imposed wide flat priors on all of the parameters, 
such that the relevant parameter space is fully sampled, and further widening these priors has no impact on the resultant constraints for parameters that are constrained by the data. For parameters not constrained by the data, e.g., $w_X(z_i)$ at $z_i>1$, we choose very wide priors to show that they are not bound by the data. 

We carry out an MCMC analysis \cite{Lewis2002} using the data, with 12 chains and a few million samples for each case, reaching excellent convergence
per Gelmann-Rubin criterion, with $R-1 < 0.005$.

Sec.\ref{sec:Xz} and Sec.\ref{sec:Xz-dz} contain our primary analysis results from using the DESI DR2 BAO measurements, combined with CMB distance priors \cite{WangPia2007,WangDai2016}. 
 Figure 1 in Sec.\ref{sec:Xz} shows our main result, which is that the $\rho_X(z)$ obtained directly from the combination of these two data sets only differs from a cosmological constant by $\sim 1 \sigma.$  Further, in the same subsection, we show that measuring $\rho_X(z)$ is preferred to measuring $w_X(z)$ by model selection via AIC and BIC, and that CMB distance priors do not bias the dark energy measurements.  
In Sec.\ref{sec:Xz-dz}, we test a variety of redshift choices for the measurements to derive the optimized $\rho_X(z)$ measurement favored by AIC and BIC.
In Sec.\ref{sec:w0wa}, we derive $(w_0,w_a)$ constraints for comparison with the work by others, using DESI DR2 BAO data combined with two different approaches for compressed CMB statistics: (1) CMB distance priors and (2) the alternative compression of CMB statistics used by the DESI team.

\subsection{Measurements of \texorpdfstring{$X(z)$}{} and  \texorpdfstring{$w_X(z)$}{} as Free Functions}
\label{sec:Xz}

Fig.1 shows the measured dark energy density $\rho_X(z)$ (left panel) and dark energy equation of state $w_X(z)$ (right panel) from our analysis in three cases:\\
\noindent
(1) DESI DR2 + Planck (magenta points and pink shaded regions\footnote{For each set of $\{X(z_i)\}$ parameters ($X(z)\equiv \rho_X(z)/\rho_X(0))$ in the MCMC chains, we obtain $X(z_j)$ at a set of densely spaced redshift values $\{z_j\}$ by cubic-splining over $\{X(z_i)\}$.
Then at each redshift $z_j$, we find the 68.3\% C.L. interval of $X(z_j)$. The shaded 68.3\% C.L. region of $X(z)$ results from connecting all of these 68.3\% C.L. intervals of $X(z_j)$. }); \\
\noindent
(2) DESI DR2 + $0.13\leq \Omega_mh^2 \leq 0.15$, $0.02\leq \Omega_bh^2 \leq 0.024$, $0.3 \leq h < 1$ (cyan points);\\
\noindent
(3) DESI DR2 + $0.11\leq \Omega_mh^2 \leq 0.17$, $0.02\leq \Omega_bh^2 \leq 0.024$, $0.3 \leq h < 1$ (blue points).

\begin{figure}
\centering
\includegraphics[width=0.495\columnwidth,clip]{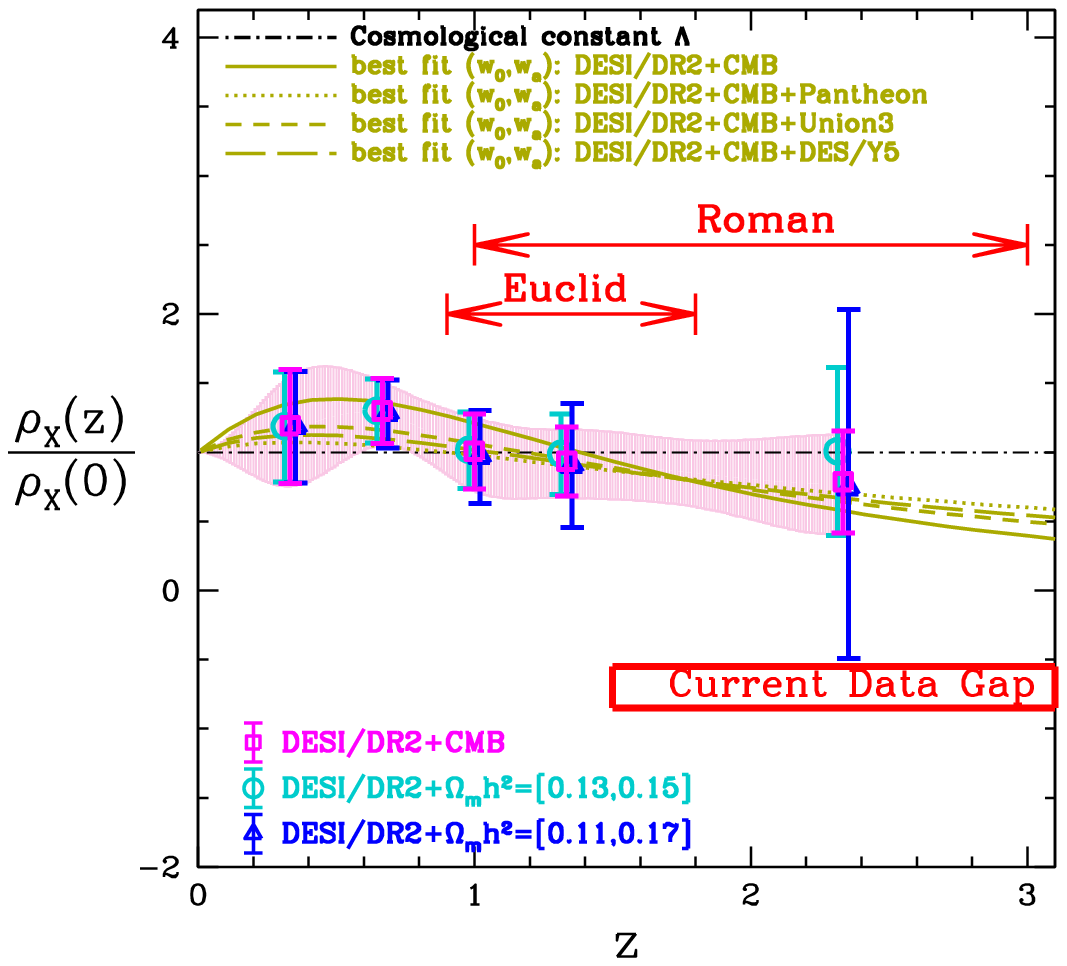}
\includegraphics[width=0.495\columnwidth,clip]{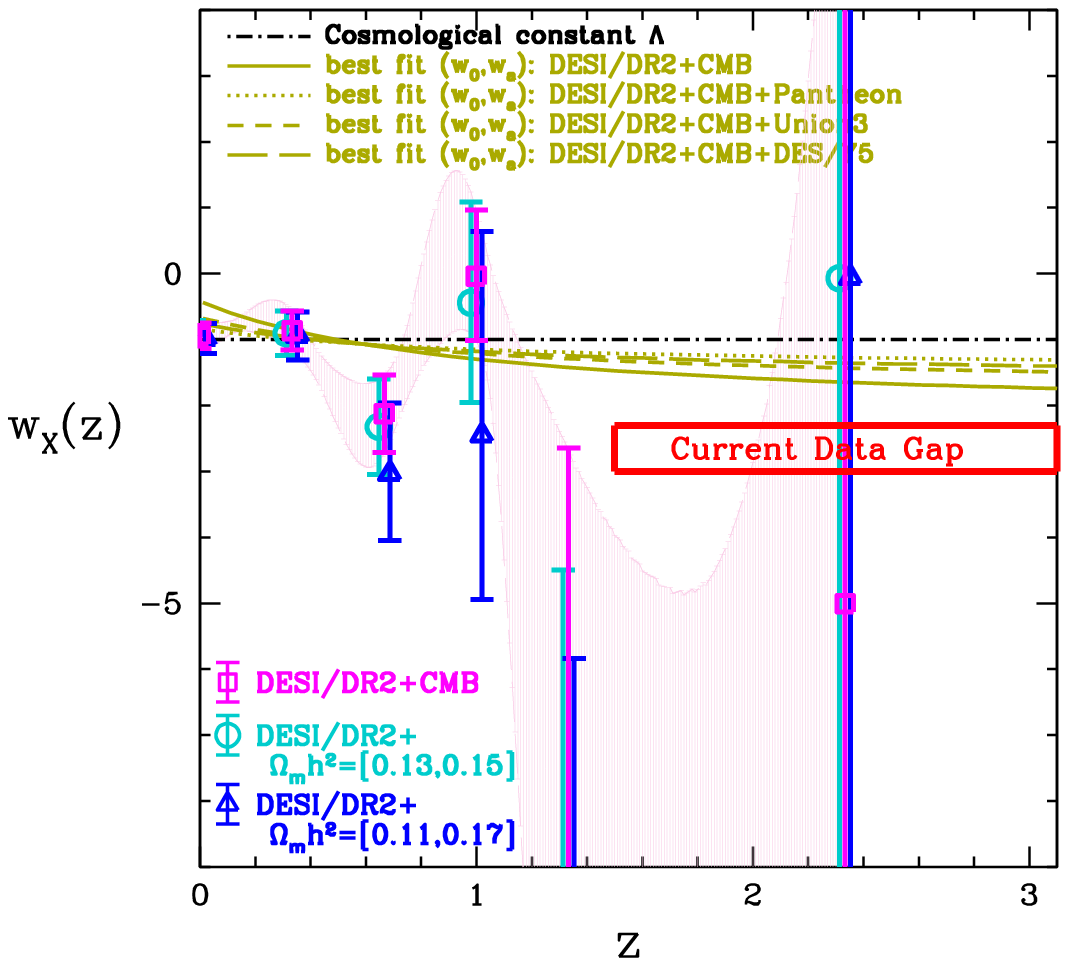}
\vspace{-0.5in}
\caption{Dark energy density $\rho_X(z)$ and dark energy equation of state $w_X(z)$ measured from DESI DR2 data. Our results, obtained with a simple nonparametric approach,  are shown in magenta for DESI DR2 data + Planck CMB distance priors from \cite{WangDai2016}, and in cyan and blue for DESI DR2 + priors on $\Omega_mh^2$ (no CMB) as labeled in the legend.  
The error bars represent the 68.3\% C.L. intervals.
The pink shaded regions correspond to the 68.3\% C.L. intervals of the interpolated $\rho_X(z)$ and $w_X(z)$ corresponding to the magenta points.
Best fit models obtained by \cite{DESI-DR2} using $w_0w_a$ parametrization are shown in olive green for data sets as labeled. For comparison with theory, the dot-dashed horizontal line is a cosmological constant.
Note that the redshift range marked ``Current Data Gap'' (there is only a single BAO measurement from DESI Ly$\alpha$ at
z = 2.33) will be covered by upcoming data from space missions Roman and Euclid, with redshift coverages as shown.
}
\label{fig:Xz-wz}
\end{figure}

For comparison with our results, best fit models obtained by \cite{DESI-DR2} using $w_0w_a$ parametrization are shown in olive green for data sets as labeled. For comparison with theory, the dot-dashed horizontal line is a cosmological constant.
There are physically motivated dark energy models, e.g., 
models inspired by decaying vacuum energy  \cite{Freese1987}
and the tracking oscillating energy model \cite{Dodelson2000} all predict very different dark energy behavior from both a cosmological constant and the $w_0w_a$ model in the redshift range marked ``Current Data Gap'' (there is only a single BAO measurement from DESI Ly$\alpha$ at
z = 2.33), which will be covered by upcoming data from the space missions Roman and Euclid, with redshift coverages as shown.

In the left panel of Fig.1, it is striking to note that the best fit $w_X(z)=w_0+w_a(1-a)$ models from \cite{DESI-DR2} all converge toward $X(z)<1$ ($X(z\gg 1)=0$) in the data gap at $z>1.5$ (there is only a single BAO measurement from DESI Ly$\alpha$ at $z=2.33$), while predicting very different $X(z)$ at lower redshifts. 
The cause for this strange high redshift behavior is: all of the best fit $(w_0,w_a)$ models from \cite{DESI-DR2} have $w_0+w_a<-1$, which drives $X(z)$ to small values at higher redshifts (see Eq.[\ref{eq:Xz_w0wa}]).
This highlights the limitation of the $(w_0,w_a)$ parametrization: with only $w_X(0)$ and $w_X'(0)$, no freedom is allowed for higher redshift behavior, thus precluding discovering how dark energy actually varies with time at higher redshifts.

When DESI data are combined with the CMB distance priors derived from Planck data, we find that $X(z)$ deviates from a cosmological constant at $\sim 1\sigma$ at $z =2/3$. 
The $X(z)$ constraints from DESI data alone but with priors on $\Omega_mh^2$ are sensitive to the choice of $\Omega_mh^2$ prior, but qualitatively consistent over a wide range of flat priors on $\Omega_m h^2$, and consistent with the DESI plus Planck results ($\sim 1\sigma$ tension with $\Lambda$CDM). 

The $w_X(z)$ constraints from DESI data alone are more sensitive to the $\Omega_mh^2$ prior, in comparison to the $\rho_X(z)$ constraints.
In the left panel of Fig. 1, one can see that the mean estimated value of $\rho_X(z)$ remains relatively unchanged for our two choices of priors, $0.13\leq \Omega_m h^2 \leq 0.15$ and the wider flat prior $0.11\leq \Omega_m h^2 \leq 0.17$.  On the other hand, in the right panel of Fig. 1, one can see that the mean estimated value of 
$w_X(z)$ is significantly shifted between the same two choices of priors on $\Omega_mh^2$.

We find that $w_X(z)$ is significantly less constrained than $\rho_X(z)$ by the same data, consistent with the finding by \cite{WangFreese2006}, since $w_X(z)$ is not constrained by data directly, but through $X(z)$ via an integral (see Eq.[\ref{eq:wX}]). This is reflected in the fact that it requires an extra parameter, $w_X(0)$, to represent $w_X(z)$ in the same redshift intervals as $X(z)$.
DESI combined with Planck data leads to a measured $w_X(z)$ that deviates from a cosmological constant at $\sim 2\sigma$ at $z=2/3$, a larger deviation than for the measured $\rho_X(z)$, but becomes unconstrained at $z>1$ (in contrast to the measured $\rho_X(z)$, see Fig.1).

Indeed, Table 1 shows the clear advantage of measuring $\rho_X(z)$ over $w_X(z)$.  The table lists the data sets and priors we used, and the resulting  $\chi^2_\nu$ ($\chi^2$ per degree of freedom), AIC, and BIC in each case for  $\rho_X(z)$ and $w_X(z)$.
Measuring $X(z)$ is preferred to measuring $w_X(z)$ at the same set of redshift values, with $\Delta\mbox{AIC}=2.61$ (moderate evidence) and $\Delta\mbox{BIC}=3.4$ (moderate evidence) in favor of measuring $X(z)$ over $w_X(z)$ from combining DESI DR2 BAO data with Planck CMB distance priors \cite{WangDai2016}.
\begin{table}
\begin{footnotesize}
\begin{tabular}{|l|l|l|l|l|l|l|}
\hline
Dark energy measurement & Data   & N$_{\rm data}$  & N$_{\rm par}$ &$\chi^2_{\nu}$ & AIC &BIC \\
\hline
$\rho_X(z_i)$, $z_i=1/3, 2/3, 1, 4/3, 2.33$ & DESI DR2 + $ 0.11\leq \Omega_mh^2 \leq 0.17$  & 13 & 8 & 0.70 & 19.52 & 24.04\\
\hline
$\rho_X(z_i)$, $z_i=1/3, 2/3, 1, 4/3, 2.33$ & DESI DR2 + $ 0.13\leq \Omega_mh^2 \leq 0.15$  & 13 & 8 & 0.71 & 19.55 & 24.07\\
\hline
$\rho_X(z_i)$, $z_i=1/3, 2/3, 1, 4/3, 2.33$ & DESI DR2 + Planck  & 16 & 8 & 0.53 & 20.20 & 26.38 \\
\hline 
\hline
$w_X(z_i)$, $z_i=0, 1/3, 2/3, 1, 4/3, 2.33$ & DESI DR2 + $ 0.11\leq \Omega_mh^2 \leq 0.17$ & 13 & 9 & 0.87 & 21.48 & 26.56 \\
\hline
$w_X(z_i)$, $z_i=0, 1/3, 2/3, 1, 4/3, 2.33$ & DESI DR2 + $ 0.13\leq \Omega_mh^2 \leq 0.15$& 13 & 9 & 0.86 & 21.43 & 26.52 \\
\hline
$w_X(z_i)$, $z_i=0, 1/3, 2/3, 1, 4/3, 2.33$ & DESI DR2 + Planck  & 16 & 9 & 0.69 & 22.81 & 29.76 \\
\hline 
\end{tabular}
\caption{AIC, BIC, and $\chi^2$ per degree of freedom for the $\rho_X(z)$ and $w_X(z)$ measurements from our MCMC analysis, for data sets and priors as shown, with redshift bins as shown. $N_{\rm data}$ is the number of data points used. $N_{\rm par}$ is the number of fitted parameters.
Flat priors $0.02\leq \Omega_bh^2 \leq 0.024$, $0.3 \leq h < 1$ are imposed in all cases.  }
\end{footnotesize}
\end{table}

\begin{figure}
\centering
\includegraphics[width=0.495\columnwidth,clip]{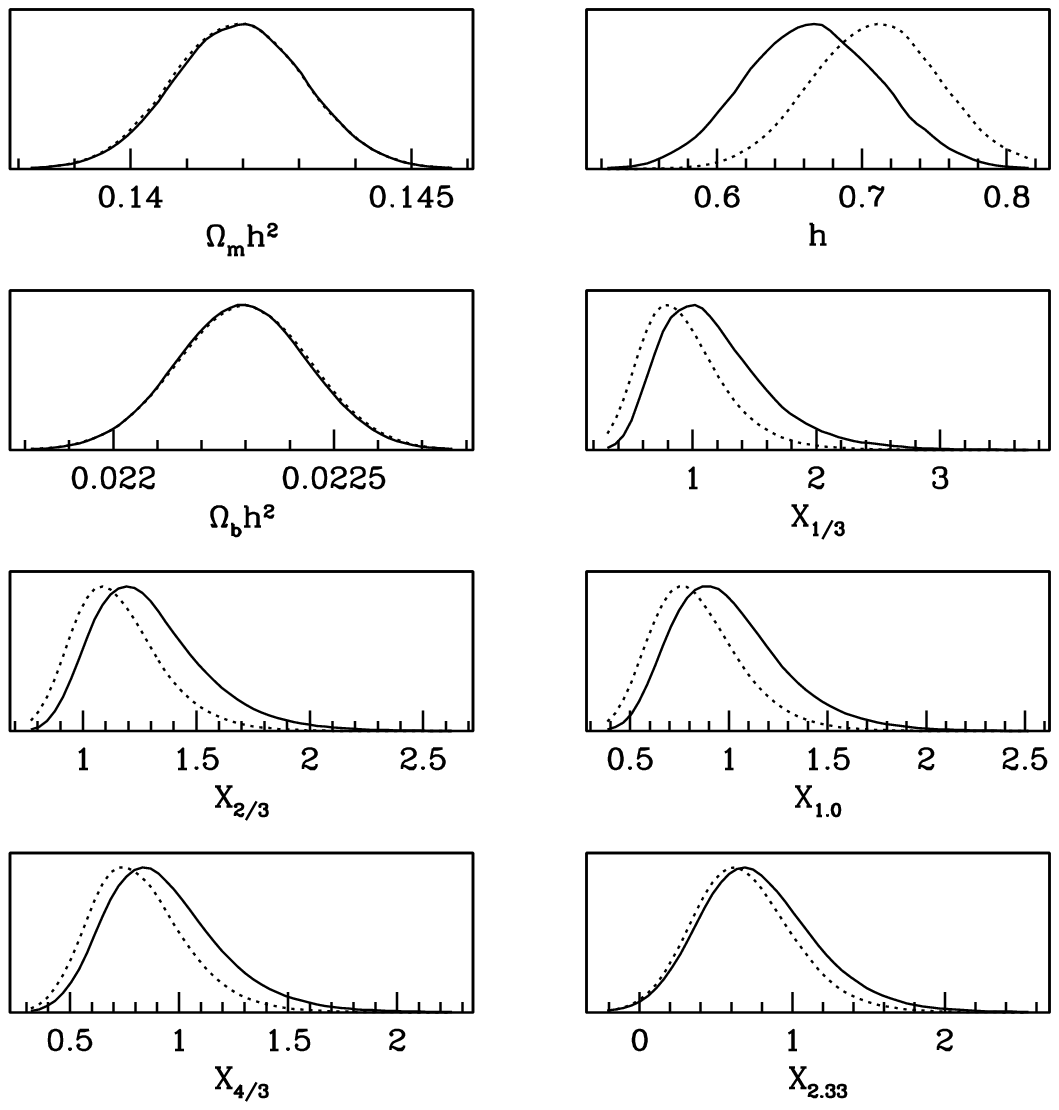}
\includegraphics[width=0.495\columnwidth,clip]{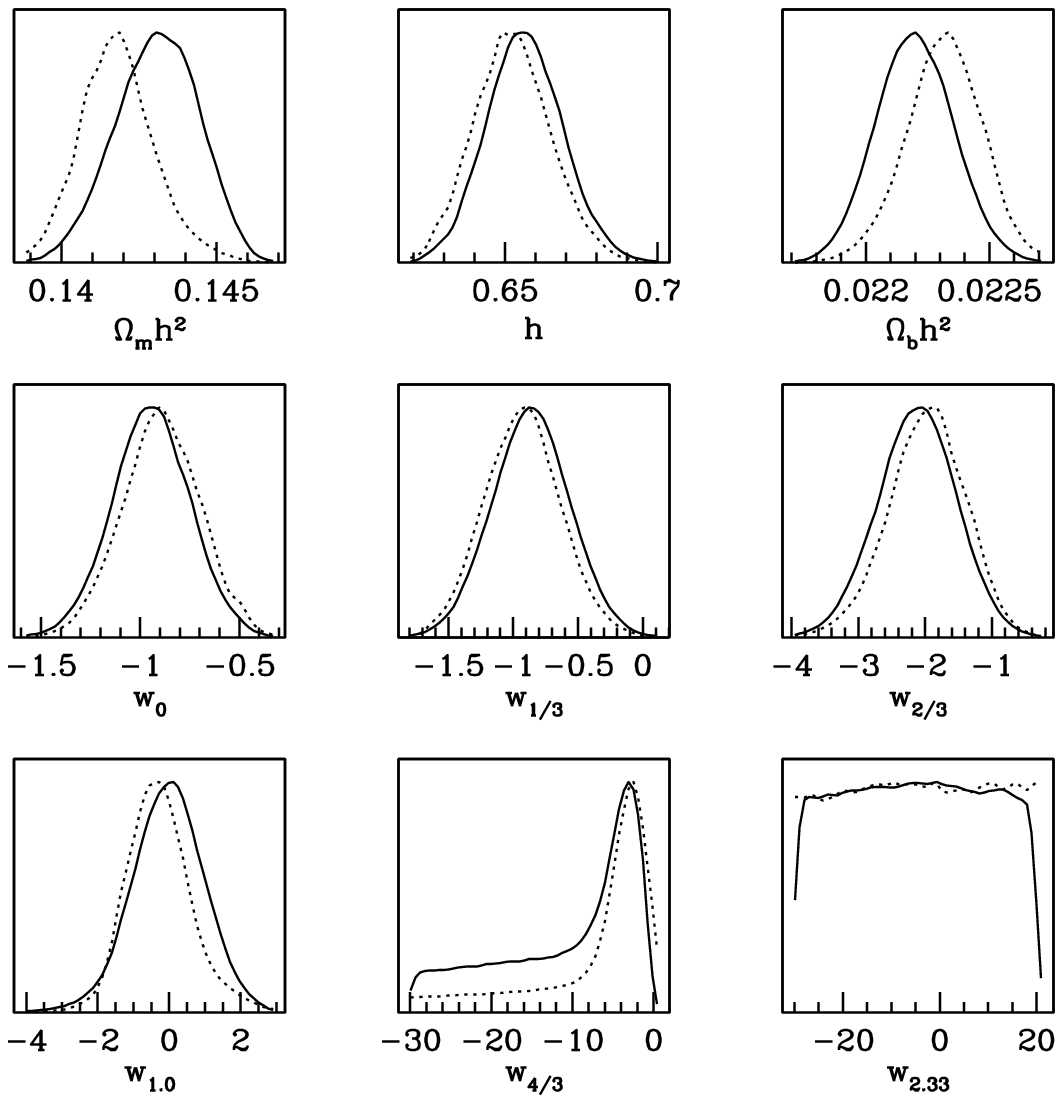}\\
\vspace{-0.1in}
\caption{Probability density distributions (pdf) for cosmological parameters $\{\Omega_m h^2, h, \Omega_b h^2\}$ and dark energy parameters measured from DESI DR2 and Planck CMB distance priors from \cite{WangDai2016}.  
The redshift values of the $X(z)\equiv \rho_X(z)/\rho_X(0)$ and $w_X(z)$ measurments are shown as subscripts for simplicity.
The solid lines show the fully marginalized posterior; the dotted lines show the relative mean likelihood of the MCMC samples.
Left panel: $\{\rho_X(1/3),\rho_X(2/3),\rho_X(1),\rho_X(4/3),\rho_X(2.33)\}$.
Right panel:
$\{w_X(0),w_X(1/3),w_X(2/3),w_X(1),w_X(4/3),w_X(2.33)\}$.}
\label{fig:pdf-DESI+Planck}
\end{figure}

\begin{figure}
%\centering
\includegraphics[width=0.495\columnwidth,clip]{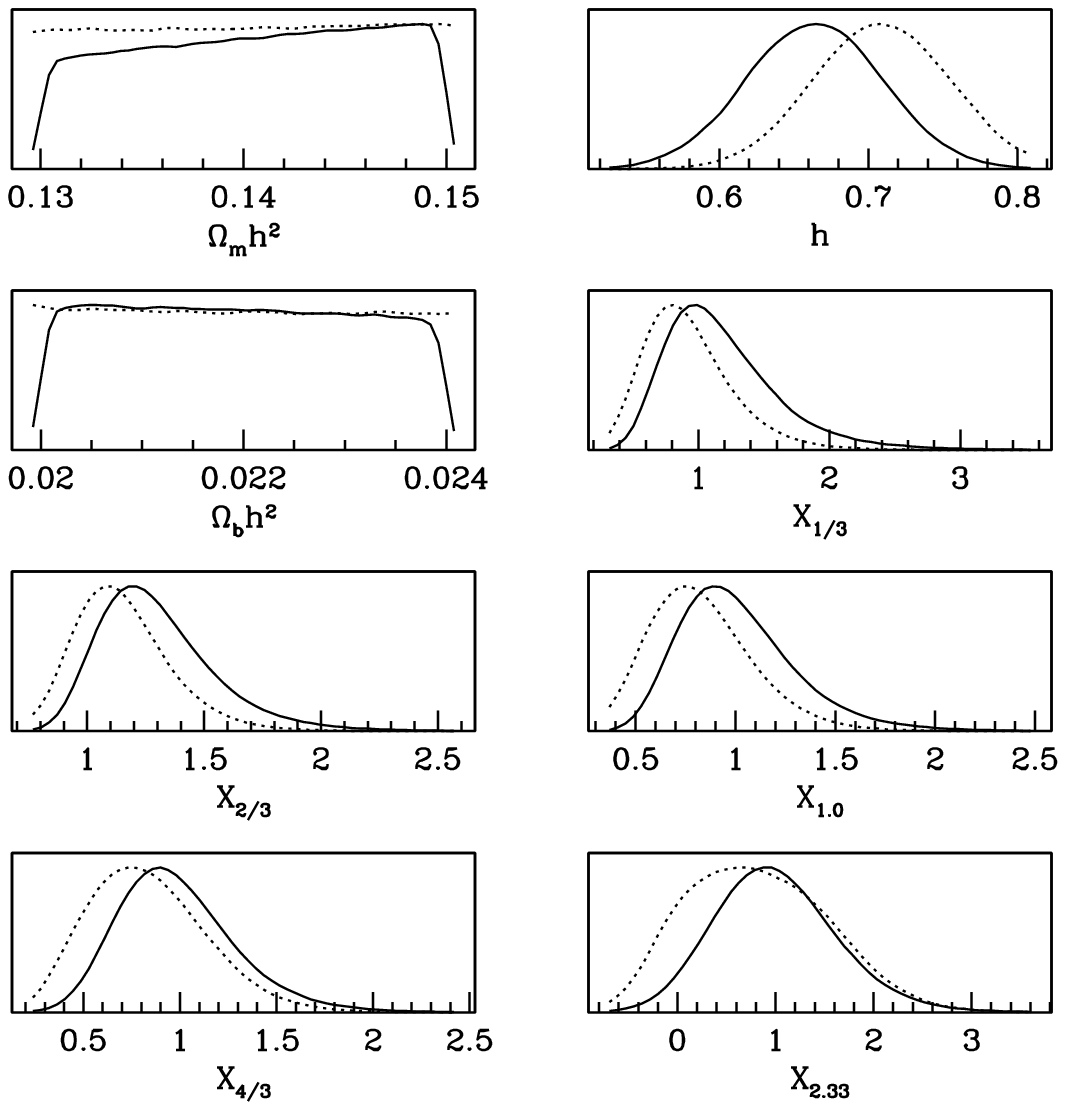}
\includegraphics[width=0.495\columnwidth,clip]{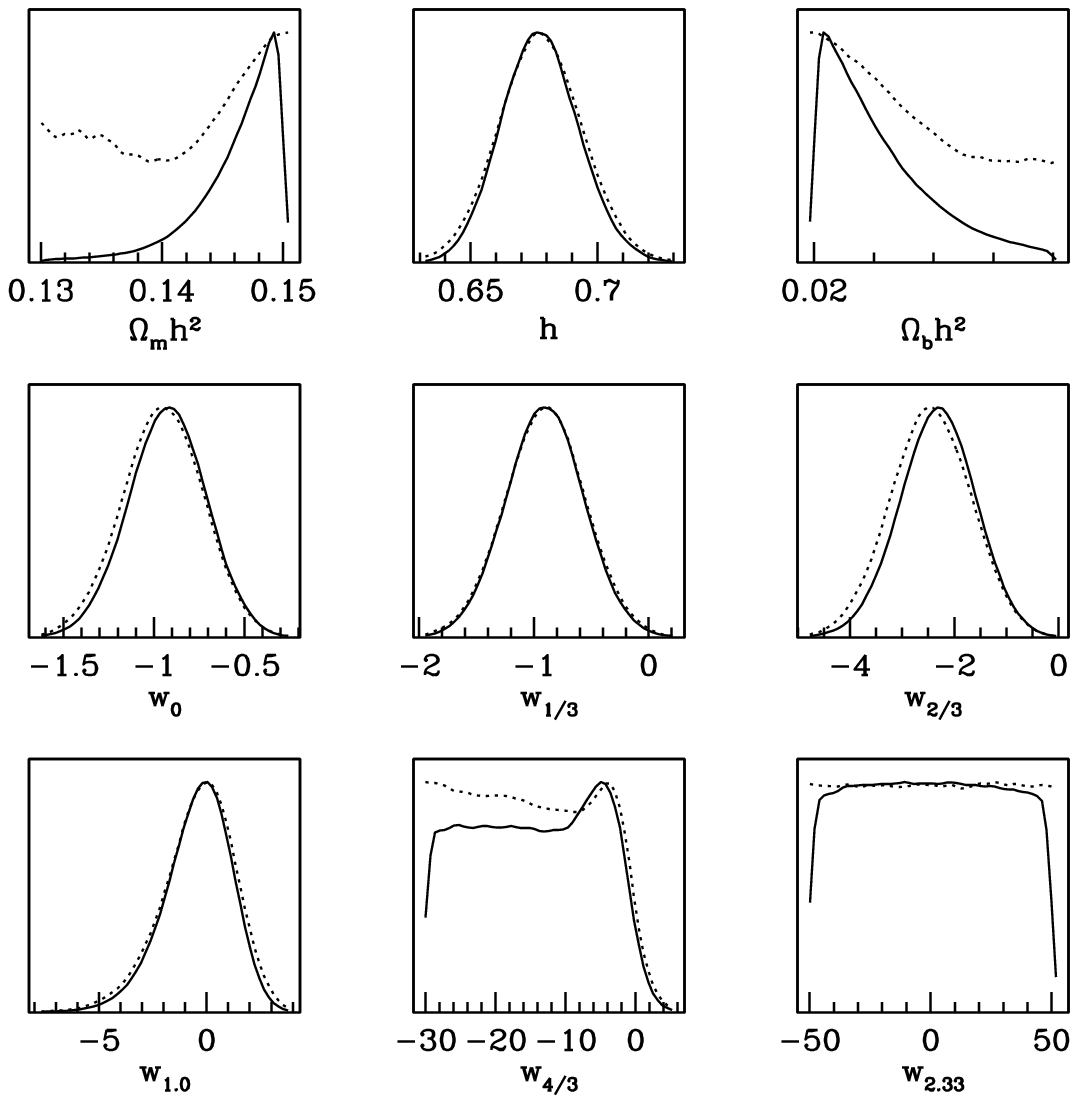}
\vspace{-0.2in}
\caption{The same as Fig.\ref{fig:pdf-DESI+Planck}, but for DESI DR2 data only with flat priors $0.13\leq \Omega_mh^2 \leq 0.15$, $0.02\leq \Omega_bh^2 \leq 0.024$, $0.3 \leq h < 1$.}
\label{fig:pdf-DESI-wideprior2}
\end{figure}

\begin{figure}
%\centering
\includegraphics[width=0.495\columnwidth,clip]{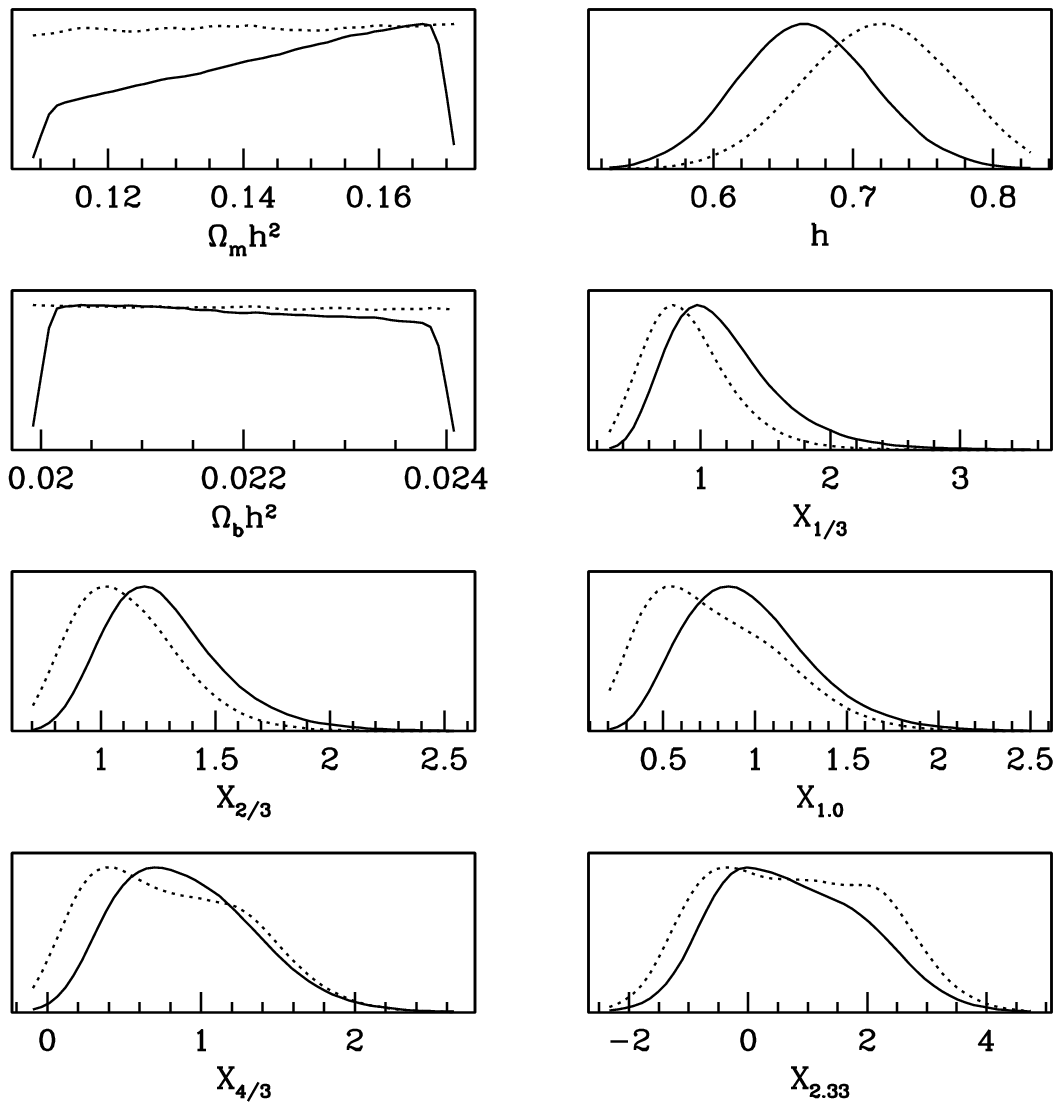}
\includegraphics[width=0.495\columnwidth,clip]{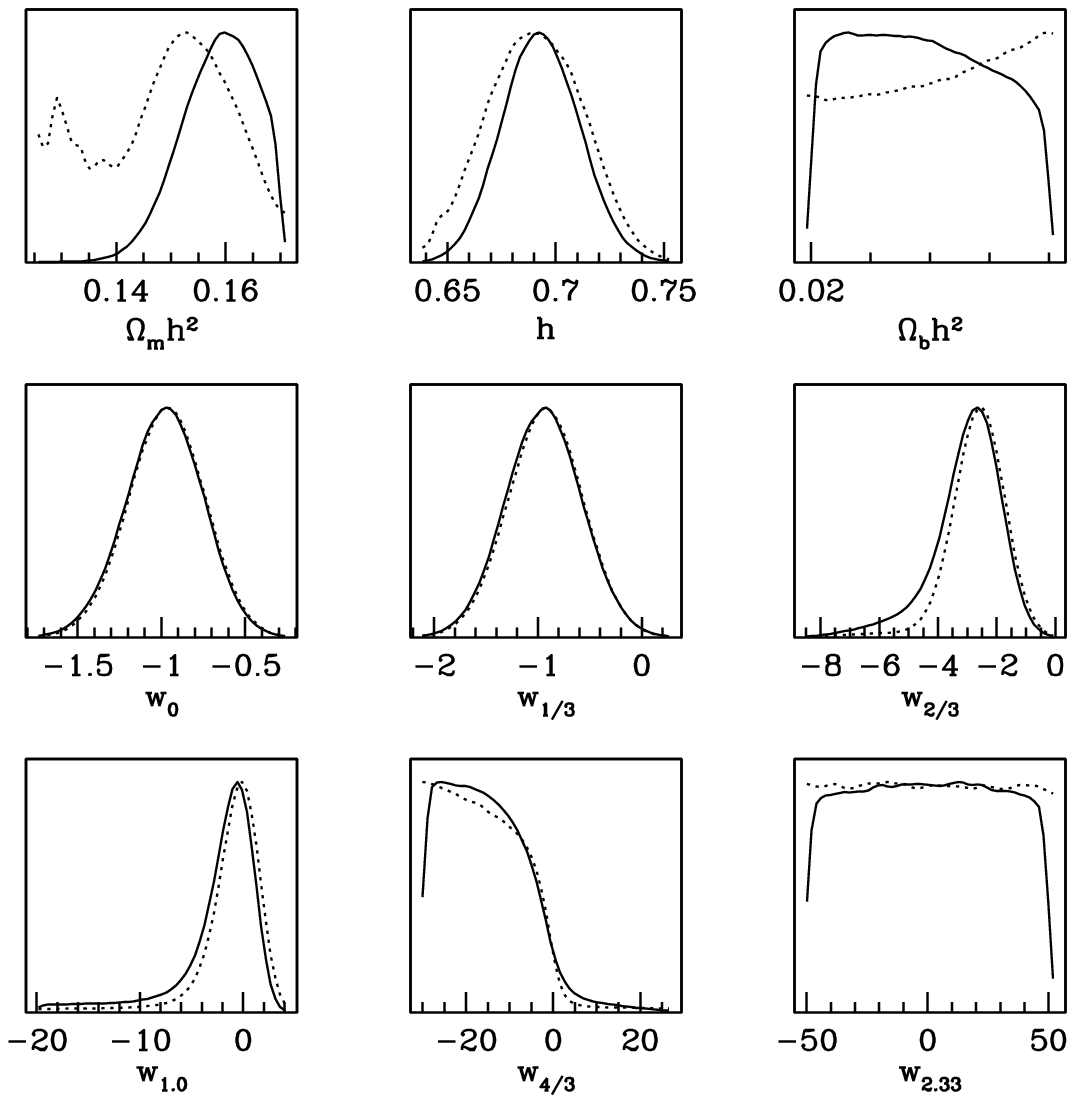}
\vspace{-0.2in}
\caption{The same as Fig.\ref{fig:pdf-DESI+Planck}, but for DESI DR2 data only with flat priors $0.11\leq \Omega_mh^2 \leq 0.17$, $0.02\leq \Omega_bh^2 \leq 0.024$, $0.3 \leq h < 1$.}
\label{fig:pdf-DESI-wideprior3}
\end{figure}

Figs.\ref{fig:pdf-DESI+Planck}-\ref{fig:pdf-DESI-wideprior3} show probability density distributions (pdf) for fully marginalized posterior (solid lines) and mean relative likelihood of the MCMC samples (dotted lines) corresponding to Fig.1.
These two distributions could differ because the marginalized posterior combines the likelihood (of the data given the parameter values) with the priors on parameters.
The priors can shift the marginalized posterior, especially if they add no information, e.g., very wide flat priors.
The mean relative likelihood gives useful complementary information to the marginalized posterior; it corresponds to the mean likelihood of the samples at each value of the parameter. In practice, it is the mean value of the likelihood projected in bins of each parameter (see Appendix C of \cite{Lewis2002}), and
a byproduct of the {\it CosmoMC} software analysis package.

Fig. 2 shows the pdf for cosmological parameters $\{\Omega_m h^2, h, \Omega_b h^2\}$ and dark energy parameters  measured from DESI DR2 and Planck CMB distance priors from \cite{WangDai2016}. Distributions are shown in the left panel for the case where we measure the energy density $X_{z_i} \equiv \rho(z_i)/\rho(z=0)$ (with redshift values given in the caption) while distributions where we measure $w(z_i)$  are shown in the right  panel.
Fig.2 shows that adding Planck distance priors to DESI DR2 data is sufficient to place constraints on dark energy, with all of the $\rho_X(z)$ parameters, and most of the $w_X(z)$ parameters are constrained by the joint data analysis.

Fig.3 shows that instead of adding Planck priors, we can impose very reasonable priors of $0.13\leq \Omega_mh^2 \leq 0.15$, $0.02\leq \Omega_bh^2 \leq 0.024$, $0.3 \leq h < 1$, and obtain results very similar to those from adding Planck distance priors as was done in Figure 2.
This indicates that our dark energy measurements are robust and not sensitive to the priors, and that Planck CMB distance priors do not introduce a bias in the dark energy measurements (as shown in \cite{WangPia2007}).
Fig.4 shows that relaxing the matter density prior to the wide range of $0.11\leq \Omega_mh^2 \leq 0.17$ still leads to meaningful constraints on dark energy.
However, comparison of the left panels (the $\rho_X(z)$ case) and right panels (the $w_X(z)$ case) in Figs.3-4 indicates that while all of the fitted parameters are well behaved\footnote{By referring to a parameter as “well behaved”, we mean that its entire probability distribution is well constrained (no cutoff or divergence at either end), and insensitive to the range of flat priors assumed for the MCMC analysis.} in the $\rho_X(z)$ case, they are not so in the $w_X(z)$ case. 

Figs.3-4 (right panels) show the fitted parameters for the $w_X(z)$ case are very sensitive to the priors imposed, with unphysical peaks in the marginalized distributions of $\Omega_mh^2$ and $\Omega_bh^2$. In contrast, in the $\rho_X(z)$ case (Fig.3-4, left panels), the marginalized distributions for $\Omega_bh^2$ are essentially flat, whereas those for $\Omega_mh^2$ have only slight tilts favoring higher $\Omega_mh^2$ 
(the rising probability density distributions are indicative of the fact that DESI prefers a higher value of $\Omega_m$).

\begin{figure}
\includegraphics[width=0.9\columnwidth,clip]{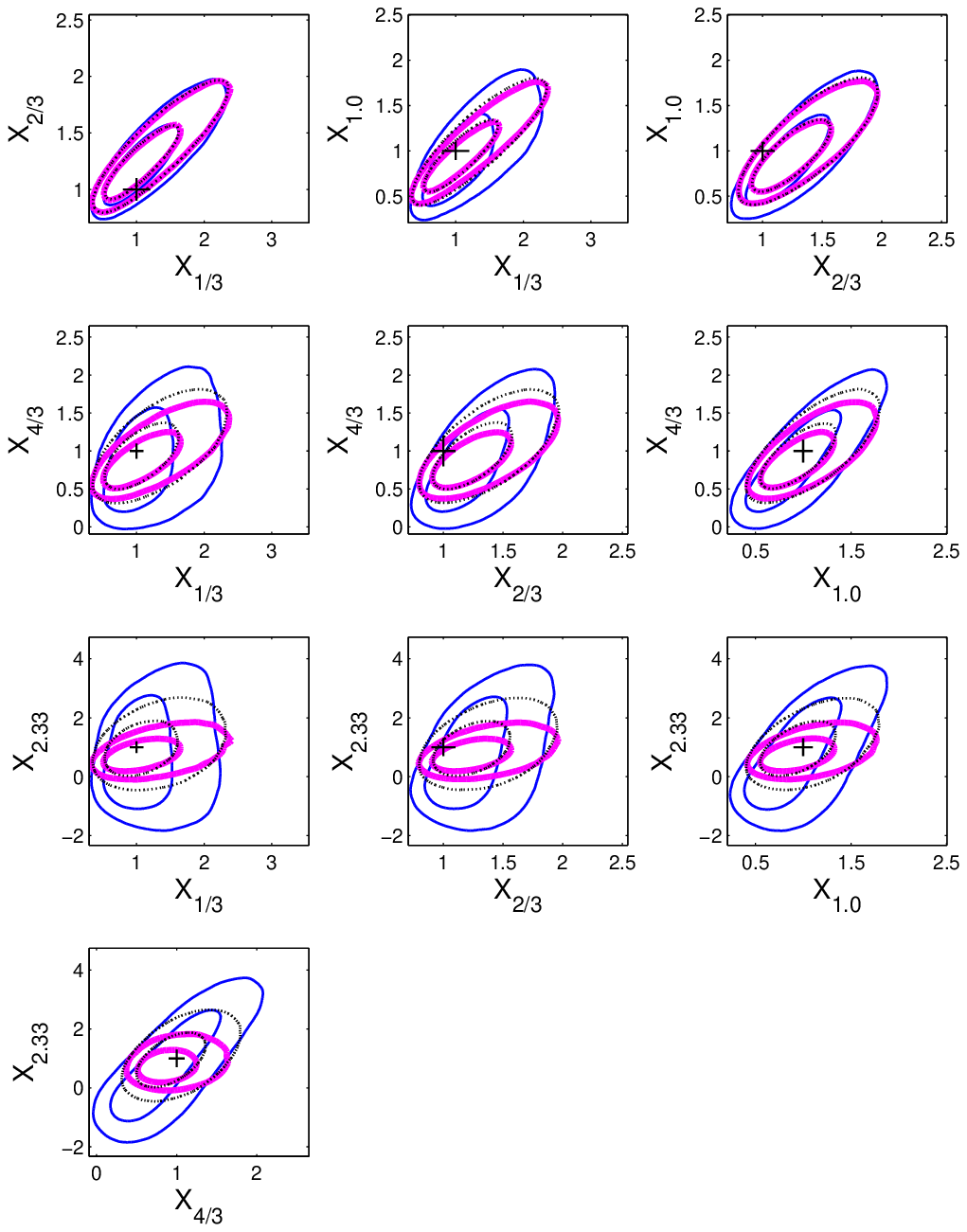}
\caption{The 68.3\% and 95.4\% joint confidence level contours of the $\rho_X(z)$ dark energy parameters corresponding to left panels in Figs.3-5, for DESI DR2 BAO plus one of the following: (1) Planck CMB priors from \cite{WangDai2016} (thick magenta solid lines), (2) flat prior $0.13\leq \Omega_m h^2 \leq 0.15$ (thin dotted black lines), and (2) flat prior $0.11\leq \Omega_m h^2 \leq 0.17$ (thin solid blue lines).
The small cross represents the cosmological constant in each contour plot.
Note that (2) is nearly identical to (1) at $z\leq 1$ (the 2D contours completely overlap in the top left and top right plots), with $\rho_X(z=2/3)$ deviating from a cosmological constant at $\sim 1\sigma$.}
\label{fig:all3Xz}
\end{figure}

\begin{figure}
\includegraphics[width=0.9\columnwidth,clip]{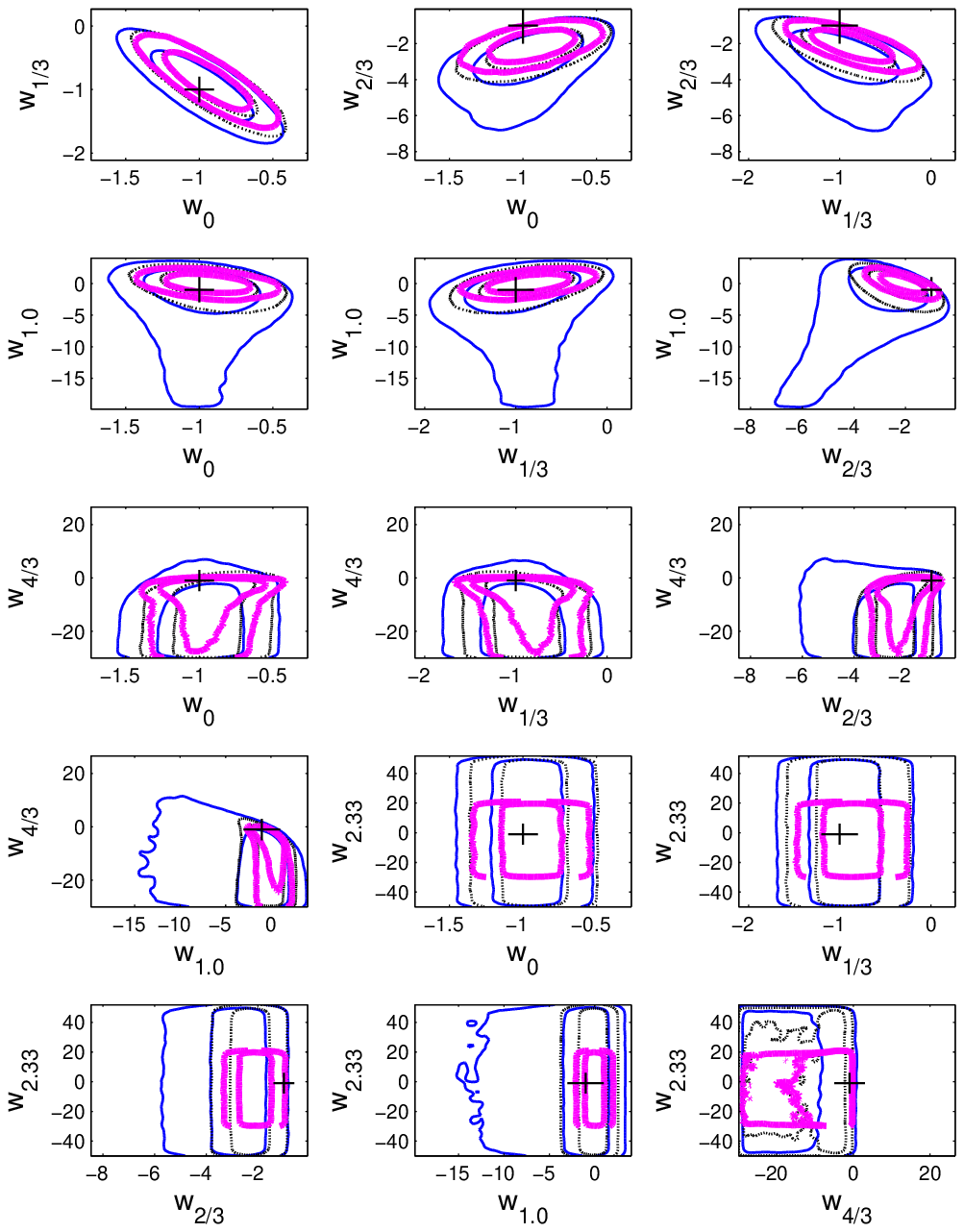}
\caption{The 68.3\% and 95.4\% joint confidence level contours of the $w_X(z)$ dark energy parameters corresponding to right panels in Figs.3-5, for DESI DR2 BAO plus one of the following: (1) Planck CMB priors from \cite{WangDai2016} (thick magenta solid lines), (2) flat prior $0.13\leq \Omega_m h^2 \leq 0.15$ (thin dotted black lines), and (3) flat prior $0.11\leq \Omega_m h^2 \leq 0.17$ (thin solid blue lines).
The small cross represents the cosmological constant in each contour plot.}
\label{fig:all3wz}
\end{figure}

Fig.\ref{fig:all3Xz} and Fig.\ref{fig:all3wz} show the 68.3\% and 95.4\% joint confidence level contours of the dark energy parameters for $\rho_X(z)$ and $w_X(z)$ respectively, corresponding to left and right panels in Figs.2-4, with the cosmological constant represented by a small cross.
Fig.\ref{fig:all3Xz} shows that compared to adding Planck distance priors, adding the flat prior $0.13\leq \Omega_m h^2 \leq 0.15$ leads to nearly identical constraints at $z\leq 1$, with $\rho_X(z=2/3)$ deviating from a cosmological constant at $\sim 1\sigma$ (see also Fig.1).
The top two rows in Fig.\ref{fig:all3wz} show that $w_X(z)$ constraints at $z\leq 1$ are highly sensitive to priors, in contrast to the $\rho_X(z)$ constraints at $z\leq 1$ (top two rows in Fig.\ref{fig:all3Xz}).
The bottom two rows in Fig.\ref{fig:all3wz} show that $w_X(z)$ is unconstrained at $z>1$ (see also Fig.1), with $w_X(z=4/3)$ only having an upper limit (no lower limit), and $w_X(z=2.33)$ completely unconstrained, as also shown in the probability density distributions for these parameters in Figs. 2-4 (right panels).

The difference between the $\rho_X(z)$ and $w_X(z)$ measurements may in part arise from the fact that the flat high $z$ cutoff at $z_{\rm max}$ has different implications for the two functions: \\
\noindent
(1) For $\rho_X(z)$, $\rho_X(z>z_{\rm max})=\rho_X(z_{\rm max})$ simply means that the dark energy density is assumed to be flat where there is no data.\\
\noindent
(2) For $w_X(z)$, $w_X(z>z_{\rm max})=w_X(z_{\rm max})$ implies $X(z)=(1+z)^{3[1+w_X(z_{\rm max})]}$ at $z>z_{\rm max}$, i.e., $\rho_X(z)$ is assumed to be a power law of $(1+z)$ where there is no data, which is a very strong assumption --- the same one implied by assuming $w_X(z)=w_0+w_az/(1+z)$
(see Eq.[\ref{eq:Xz_w0wa}]).

If we want to make the equivalent weak assumption of $\rho_X(z>z_{\rm max})=\rho_X(z_{\rm max})$ for $w_X(z>z_{\rm max})$, we need to impose $w_X(z>z_{\rm max})=-1$, which leads to a discontinuity at $z=z_{\rm max}$, thus requiring at least one more additional parameter to smooth it at $z=z_{\rm max}$.
This is another implicit disadvantage of measuring $w_X(z)$ from data.

Comparing the left panel ($X(z)$) and right panel ($w_X(z)$) in Figs.\ref{fig:pdf-DESI+Planck}-\ref{fig:all3wz}, it is clear that all of the $X(z)$ parameters are well constrained by the data, while $w_X(1.33)$ only has an upper limit (no lower limit), and $w_X(2.33)$ is unconstrained by the data.
This indicates that the $\{X(z_i)\}$ are the preferred physical parameters to measure from the data, instead of the $\{w_X(z_i)\}$.
This is as expected, since $X(z)$ enters the theoretical prediction in a straightforward manner (see Eq.[\ref{eq:basics}]), while $w_X(z)$ enters via an integral through $X(z)$ (see Eq.[\ref{eq:wX}]).

Figure 1 illustrates another advantage of measuring $X(z)$ over $w_X(z)$. 
The $X(z)$ mean value is well behaved for all $z$ 
as $z$ increases, whereas the mean value of $w_X(z)$ leaps to large negative values with large error bars at $z\geq 4/3$, due to prior volume effects\footnote{By ``prior volume effects", we refer to the fact that the measured $w_X(z)$ values are sensitive to the range of the flat priors imposed, since they are not bound from below, and could be as negative as one wants them to be.}.
As $w_X(z)$ becomes more and more negative in Eq.(2), the role of dark energy in cosmology becomes utterly negligible;
hence a large volume of negative $w_X(z)$ is equivalent in the MCMC chains (see the one-sided pdf for $w_X(4/3)$ and the flat pdf for $w_X(2.33)$ in Figs.2-4), pulling the probability distributions of $w_X(4/3)$ and $w_X(2.33)$ to more negative values and giving rise to large error bars.

Table \ref{tab:margestats_planck15} shows the mean, standard deviation, and 68\% C.L. ranges for the $X(z_i)$ measured from DESI DR2 data combined with Planck 2015 distance priors with $m_\nu=0.06\,$eV \cite{WangDai2016} (plotted in Fig.1), and Planck 2015 and 2018 distance priors allowing $m_\nu$ to vary \cite{Zhai2020}.
As expected, allowing neutrino mass to vary gives less stringent constraints, but the difference is very small except for $X(2.33)$ ($\sim$ 20\% larger when varying $m_\nu$).
The Planck 2015 and 2018 distance priors allowing $m_\nu$ to vary \cite{Zhai2020} give very similar results.
The three different Planck distance priors give nearly identical constraints on the deviations from a cosmological constant: $\sim 1\sigma$ at $z\sim 2/3$.
This comparison indicates that the Planck 2015 distance priors from \cite{WangDai2016} provides correct CMB constraints for fixing $m_\nu=0.06\,$eV, and that our measurement of $X(z)$ is robust and model-independent.

\begin{table}
\begin{footnotesize}
\begin{tabular}{|l|lll||lll||lll|}
\hline
& Planck & (2015), & fixing $m_\nu$ &Planck & (2015), & varying $m_\nu$ &   Planck & (2018), & varying $m_\nu$\\
\hline
& mean & $\sigma$ & 68.3\% C.L. & mean & $\sigma$ & 68.3\% C.L. 
& mean & $\sigma$ & 68.3\% C.L.  \\
\hline
   $ X_{1/3}$ & 1.1926 & 0.4545 & (0.7747, 1.5996) & 
   1.1853 & 0.4576 & (0.7719, 1.5840) & 
   1.1833 & 0.4511 & (0.7784, 1.5768) \\ 
    $X_{2/3}$ & 1.2989 & 0.2545 & (1.0619, 1.5328) & 
    1.2854 &  0.2545 &  (1.0518, 1.5125) & 
    1.2817 &  0.2513 &  (1.0517, 1.5072) \\ 
    $X_{1.0}$ & 1.0093 & 0.2934 & (0.7344, 1.2797) & 
    0.9804 &  0.2895 &  (0.7125, 1.2403) &
    0.9727 &  0.2846 &  (0.7099, 1.2305) \\
    $X_{4/3}$ & 0.9355 & 0.2685 & (0.6835, 1.1849) & 
    0.9212 &  0.2763 & (0.6620, 1.1755) &
    0.9101 &  0.2730 &  (0.6541, 1.1618) \\ 
    $X_{2.33}$& 0.7869 & 0.3905 & (0.4170, 1.1553) & 
    0.7896 & 0.4796 & (0.3349, 1.2406) &
    0.7565 &  0.4716 &  (0.3090, 1.2028)\\ 
\hline 
\end{tabular}
\caption{$X(z_i)$ measured from DESI DR2 data combined with Planck 2015 distance priors from \cite{WangDai2016} with $m_\nu=0.06\,$eV, and Planck 2015 and 2018 distance priors from \cite{Zhai2020} allowing $m_\nu$ to vary.}
\label{tab:margestats_planck15}
\end{footnotesize}
\end{table}

\subsection{Varying the redshift choices to optimize the  \texorpdfstring{$\rho_X(z)$}{TEXT} measurement}
\label{sec:Xz-dz}

\begin{figure}
\centering
\includegraphics[width=\columnwidth,clip]{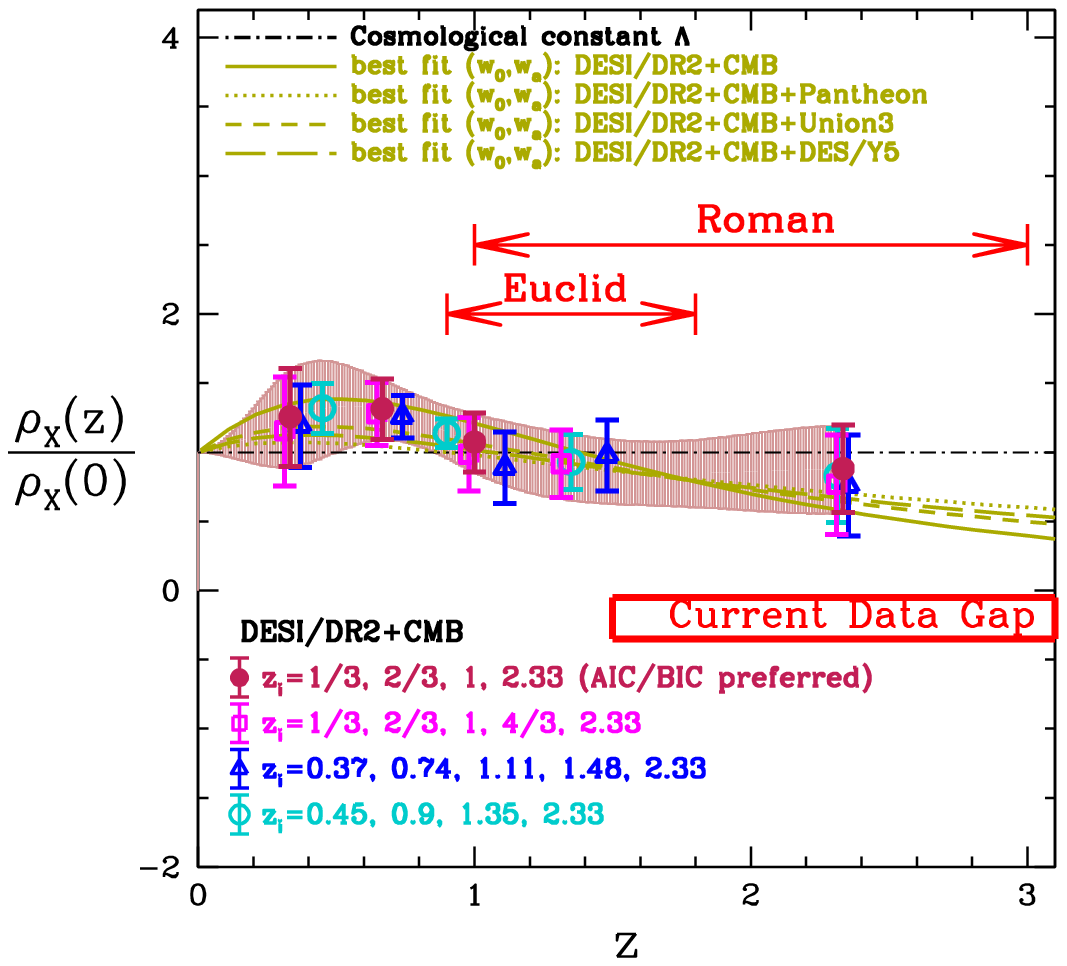}
\vspace{-0.8in}
\caption{Dark energy density $\rho_X(z)$ obtained with a simple nonparametric approach, for DESI DR2 BAO data + Planck CMB distance priors from \cite{WangDai2016}, with four different choices of dividing up the redshift range for the $\rho_X(z)$ measurement, as indicated in the figure legend.
The reddish shaded region indicates the 68.3\% confidence level contours of $\rho_X(z)$ corresponding to the AIC/BIC favored choice with d$z=1/3$ (dark red points),
which is the same as the case in Fig.1 (reproduced here as magenta points) but with the $z=4/3$ point removed. The $\rho_X(z)$ parametrizations with d$z=0.37$ (blue points) and d$z=0.45$ (cyan points) lead to results that are consistent within 1$\sigma$ with the fiducial case, with the larger ``bin-size" resulting in smaller statistical errors at $z<1$, where the DESI DR2 BAO measurements are most precise.
Thus our $\rho_X(z)$ measurement is insensitive to the choice of redshifts used for the interpolation, and deviate from a cosmological constant at 1-2$\sigma$ over the same redshift range where DESI DR2 BAO measurements deviate from a cosmological constant.
Our method is effective in accurately translating the BAO measurements into dark energy density measurements.}
\label{fig:Xz-dz}
\end{figure}

In order to demonstrate the robustness of the $ \rho_X(z)$ measurement shown in Fig.1, we test their dependence on different choices of redshift at which we measure $X(z)\equiv \rho_X(z)/\rho_X(0)$.  Note that we are not actually binning in redshift, but choosing a set of redshift values $\{z_i\}$ for interpolating $X(z)$, with $\{X(z_i)\}$
estimated as free parameters from data.
Fig.\ref{fig:Xz-dz} shows the results from our MCMC analysis of DESI DR2 BAO data combined with Planck CMB priors for four different choices for dividing up the redshift range for the $X(z)\equiv \rho_X(z)/\rho_X(0)$ measurement (each of these will be considered a ``model'' for computing AIC/BIC). 
We parametrize $X(z)$ with its values at $\{z_i\}$, \{$X(z_i)$\}, with the four different choices listed below (the redshift ``bin-size'' d$z$ is defined to be the spacing between the evenly spaced redshift values):\\
\noindent
(1) $\{z_i\}=\{1/3, 2/3, 1, 4/3, 2.33\}$ (d$z=1/3$), the fiducial case shown in Fig.1. \\
\noindent
(2) $\{z_i\}=\{0.37, 0.74, 1.11, 1.48, 2.33\}$ (d$z=0.37$), with the same number of parameters as in (1).\\
\noindent
(3) $\{z_i\}=\{0.45, 0.90, 1.35, 2.33\}$ (d$z=0.45$), with one less parameter than in (1), i.e., 4 redshifts instead of 5.\\
\noindent
(4) $\{z_i\}=\{1/3, 2/3, 1, 2.33\}$ (d$z=1/3$), the same as (1) but with one less parameter, i.e., with $z=4/3$ removed. 

All of our four model choices for $\{z_i\}$ have evenly spaced redshift values except for the final two redshifts, since in all cases, the Lyman-$\alpha$ measurements are at  $z_{\mbox{max}}=2.33$.
The DESI DR2 BAO measurements are only made at $z=0.295, 0.510, 0.706, 0.934, 1.321, 1.484, 2.33$, which constrains our possible choices for $\{z_i\}$.

Table 2 shows the AIC and BIC values for our
$\rho_X(z_i)$ measurements with four different choices for $\{z_i\}$, for DESI DR2 BAO measurements combined with Planck CMB distance priors from \cite{WangDai2016}.
All of the models have $\Delta\mbox{AIC}<2$ and $\Delta\mbox{BIC}<2$, thus they are all essentially equivalent, with a slight preference for the fourth case, i.e., 
our fiducial model with the $X(z=4/3)$ removed. This is not surprising, since from Fig.1, it is clear that $X(z)$ is consistent with $X(z)=1$ at $z>1$, thus $X(z=4/3)$ provides negligible additional information.

\begin{table}
\begin{footnotesize}
\begin{tabular}{|l|l|l|l|l|l|}\hline
$\{z_i\}$ used in $\rho_X(z_i)$ & N$_{\mbox{par}}$ & $\chi^2$ & $\chi^2_\nu$ & AIC & BIC \\
\hline
\{1/3, 2/3, 1, 4/3, 2.33\} & 8 & 4.200 & 0.525 & 20.20 & 26.38 \\
\hline
\{0.37, 0.74, 1.11, 1.48, 2.33\} & 8 & 4.800  & 0.600 & 20.80 & 26.98\\
\hline
\{0.45, 0.90, 1.35, 2.33\} & 7 & 6.234 & 0.693 & 20.23 & 25.64 \\
\hline
{\bf \{1/3, 2/3, 1, 2.33\}} & {\bf 7} & {\bf 5.340} & {\bf 0.593} & {\bf 19.34} & {\bf 24.75}\\
\hline
\end{tabular}
\caption{Comparison of $\rho_X(z_i)$ with four different choices for $\{z_i\}$, for DESI DR2 BAO measurements combined with Planck CMB distance priors from \cite{WangDai2016}, N$_{\mbox{data}}$=16.
The AIC and BIC favored parametrization is marked in bold face.}
\label{tab:z-bins-opt}
\end{footnotesize}
\end{table}

\begin{table}
\begin{footnotesize}
\begin{tabular}{|l|l||l|l||l|l||l|l|}\hline
$\{z_i\}$ & $X(z_i)$ &$\{z_i\}$ &
$X(z_i)$ & $\{z_i\}$ & $X(z_i)$ &$\mathbf{\{z_i\}}$ &
$X(z_i)$ \\
\hline\hline
   $ {1/3}$ & 1.193 (0.775, 1.600) & ${0.37}$ & 1.193 (0.8882 1.487)
   & ${0.45}$ & 1.318 (1.137, 1.499)
   & {\bf 1/3} & {\bf 1.25 (0.897, 1.608)}
\\ \hline
    ${2/3}$ & 1.299 (1.062, 1.533) & ${0.74}$ &  1.259 (1.105, 1.410)
     &${0.9}$& 1.138 (1.036, 1.240)
     & {\bf 2/3} & {\bf 1.315 (1.093, 1.531)} 
\\ \hline
    1.0 & 1.009 (0.734, 1.280) & ${1.11}$ & 0.891 (0.631, 1.145)
     & ${1.35}$ & 0.929 (0.730, 1.128) 
     & {\bf 1.0} & {\bf 1.074 (0.859, 1.286)} 
\\ \hline
    ${4/3}$ & 0.936 (0.684, 1.185) & ${1.48}$ & 0.977 (0.721, 1.233)
     & $ {2.33}$ & 0.830 (0.490, 1.171) & 
    {\bf 2.33} & {\bf 0.882 (0.563, 1.197)}
\\ \hline
    ${2.33}$& 0.787 (0.417, 1.155) & ${2.33}$ & 0.760 (0.395, 1.124)
   & & & &
   \\ 
\hline
\end{tabular}
\caption{$X(z_i)$ measurements corresponding to Table \ref{tab:z-bins-opt}, with mean values and 68\% C.L. ranges for each parameter.
The AIC and BIC favored parametrization is marked in bold face.}
\end{footnotesize}
\end{table}

The reddish shaded region in Fig.\ref{fig:Xz-dz} shows the 68.3\% confidence level contour of $X(z)$ corresponding to the AIC and BIC preferred parametrization with d$z=1/3$ and omitting $X(z=4/3)$ (dark red filled points); it's very similar to the d$z=1/3$ with $X(z=4/3)$ parametrization (the fidicial case of Fig 1 reproduced here as magenta points), with slightly smaller error bars for $X(z_i)$ due to the smaller number of parameters. The $X(z)$ parametrizations with d$z=0.37$ (blue points) and d$z= 0.45$ (cyan points) lead to results that are consistent within 1$\sigma$ with the AIC/BIC preferred parametrization, with the larger ``bin-size" naturally resulting in smaller statistical errors at $z<1$, where the DESI DR2 BAO measurements are most precise.
Our $X(z)$ constraints are consistent from the four different sets of choices of redshift values for the measurements. Shifting the redshift values for the $X(z)$ measurements consistently finds  1-2$\sigma$ deviation from a cosmological constant over the redshift range of $0.4-0.9$, corresponding precisely to where DESI DR2 BAO measurements deviate from a cosmological constant (see Fig.6 of \cite{DESI-DR2}), and with about the same statistical significance. 
This indicates that our method is effective in accurately translating the BAO measurements into dark energy density measurements.

We also tried dividing the redshift range more finely, e.g., with d$z=0.3$, in measuring $X(z)$, and found that the MCMC chains have poor convergence, which indicates that DESI DR2 BAO measurements do not favor such choices. This is as expected, since there is only one DESI BAO measurement at $z\leq 0.3$ (BGS at $z=0.295$).

If the ``bin-size" is too large, the detailed behavior of $X(z)$ is washed out. If the ``bin-size" is too small, noise dominates, and no meaningful constraints on $X(z)$ can be derived. Our fiducial case of d$z=1/3$ is optimal since it is the smallest ``bin-size'' that gives convergent constraints in the MCMC analysis, and captures the detailed behavior of $X(z)$ as a function of $z$ (see Fig.\ref{fig:Xz-dz}).

\begin{figure}
\centering
\includegraphics[width=0.49\columnwidth,clip]{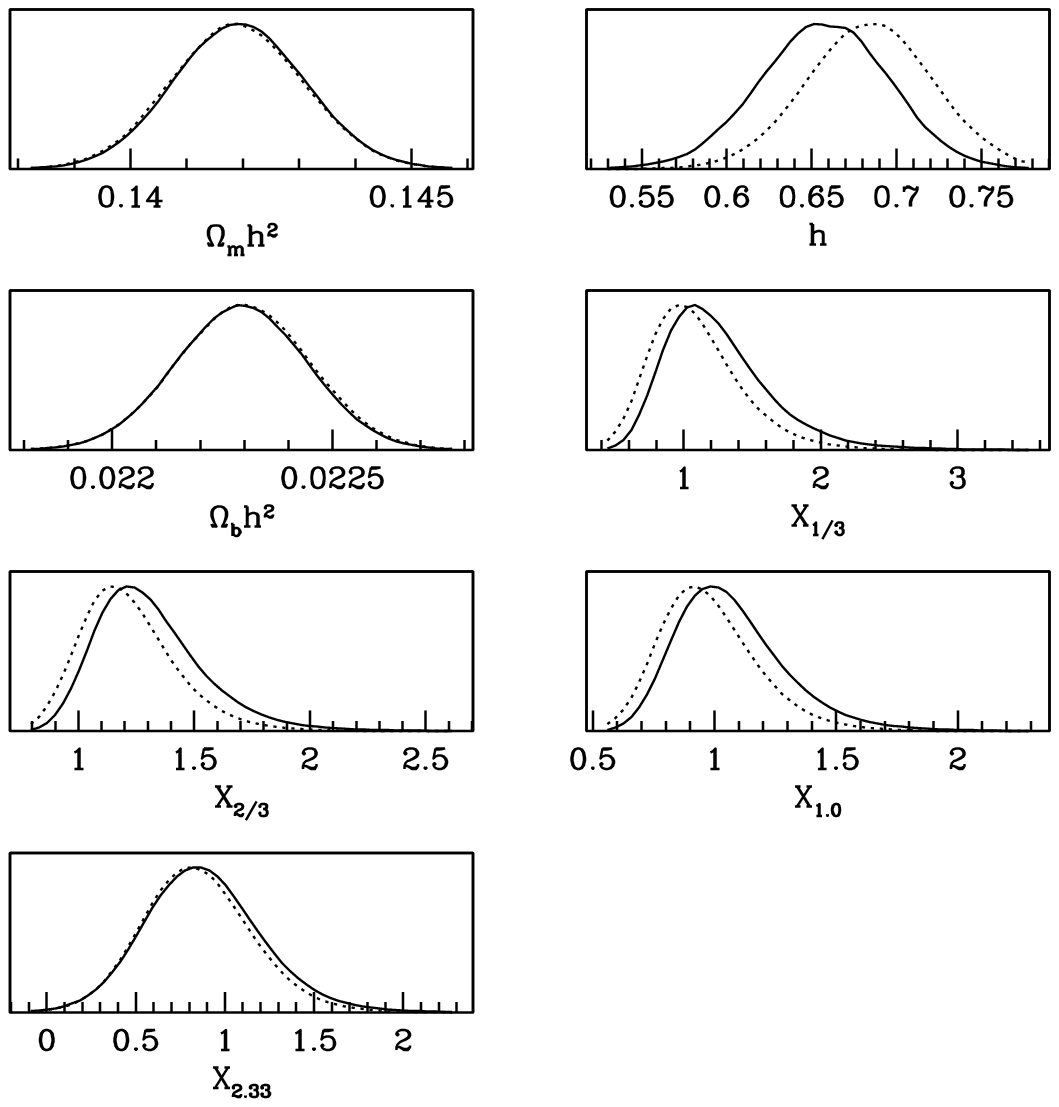}\hspace{0.05in}
\includegraphics[width=0.49\columnwidth,clip]{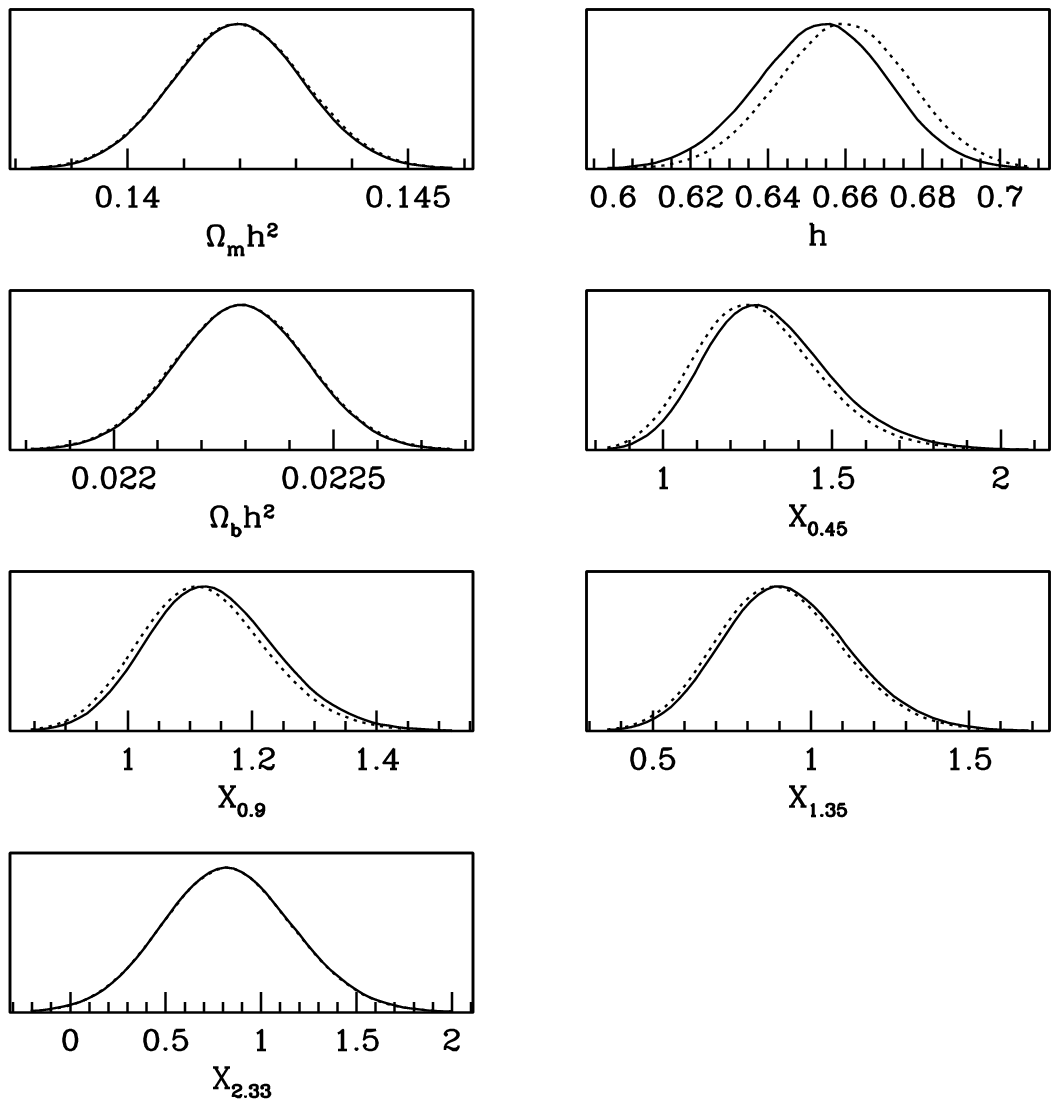}\\
%\vspace{-0.2in}
\includegraphics[width=0.45\columnwidth,clip]{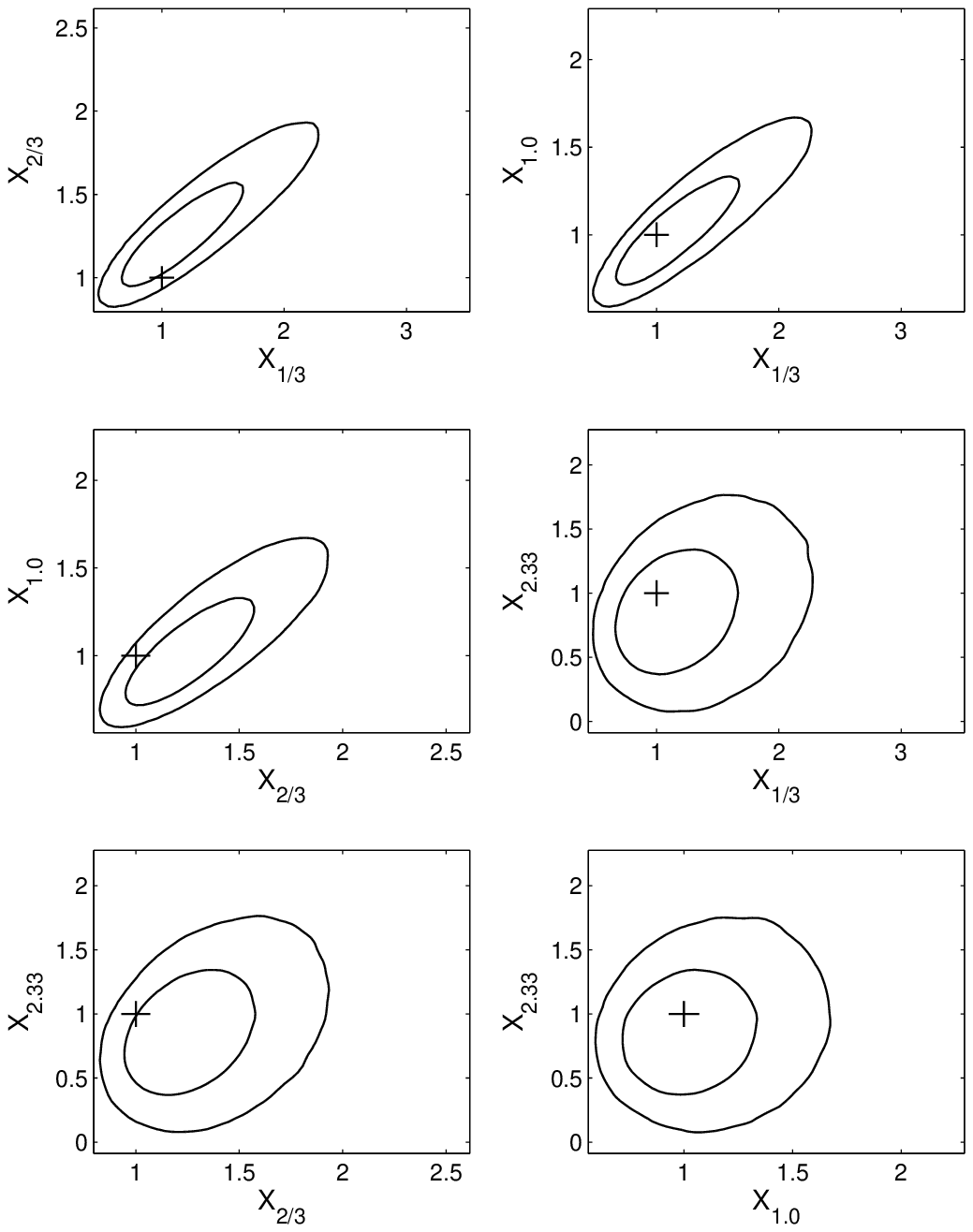}\hspace{0.3in}
\includegraphics[width=0.45\columnwidth,clip]{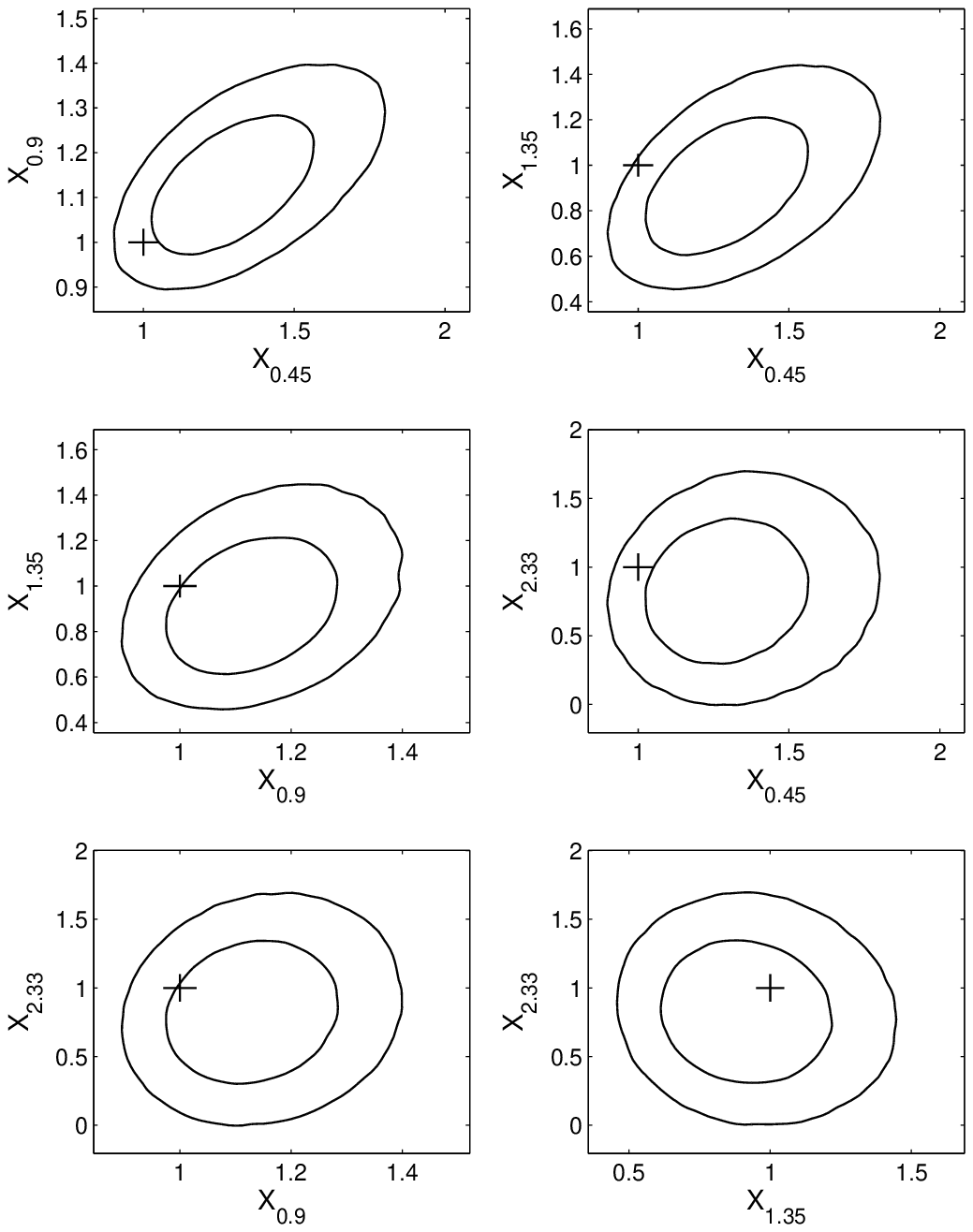}\\
\caption{Same as Fig.2 and Fig.5, but for measuring $\rho_X(z)$ using
 DESI DR2 BAO and Planck CMB distance priors from \cite{WangDai2016} at
$\{z_i\}=\{1/3, 2/3, 1, 2.33\}$ (left panels), and $\{z_i\}=\{0.45, 0.9, 1.35, 2.33\}$ (right panels).
Note that there are four redshifts in each case rather than the five earlier in the paper; the left panel is identical to the earlier fiducial case but with $z=4/3$ removed.
Top panels: Probability density distributions for cosmological parameters $\{\Omega_m h^2, h, \Omega_b h^2\}$ and dark energy parameters.
The solid lines show the fully marginalized posterior; the dotted lines show the relative mean likelihood of the MCMC samples.
Bottom panels: the corresponding 68.3\% and 95.4\% joint confidence level contours of the dark energy parameters, with 
the small cross representing the cosmological constant.}
\label{fig:alt-Xz4}
\end{figure}

\begin{figure}
\centering
\includegraphics[width=0.495\columnwidth,clip]{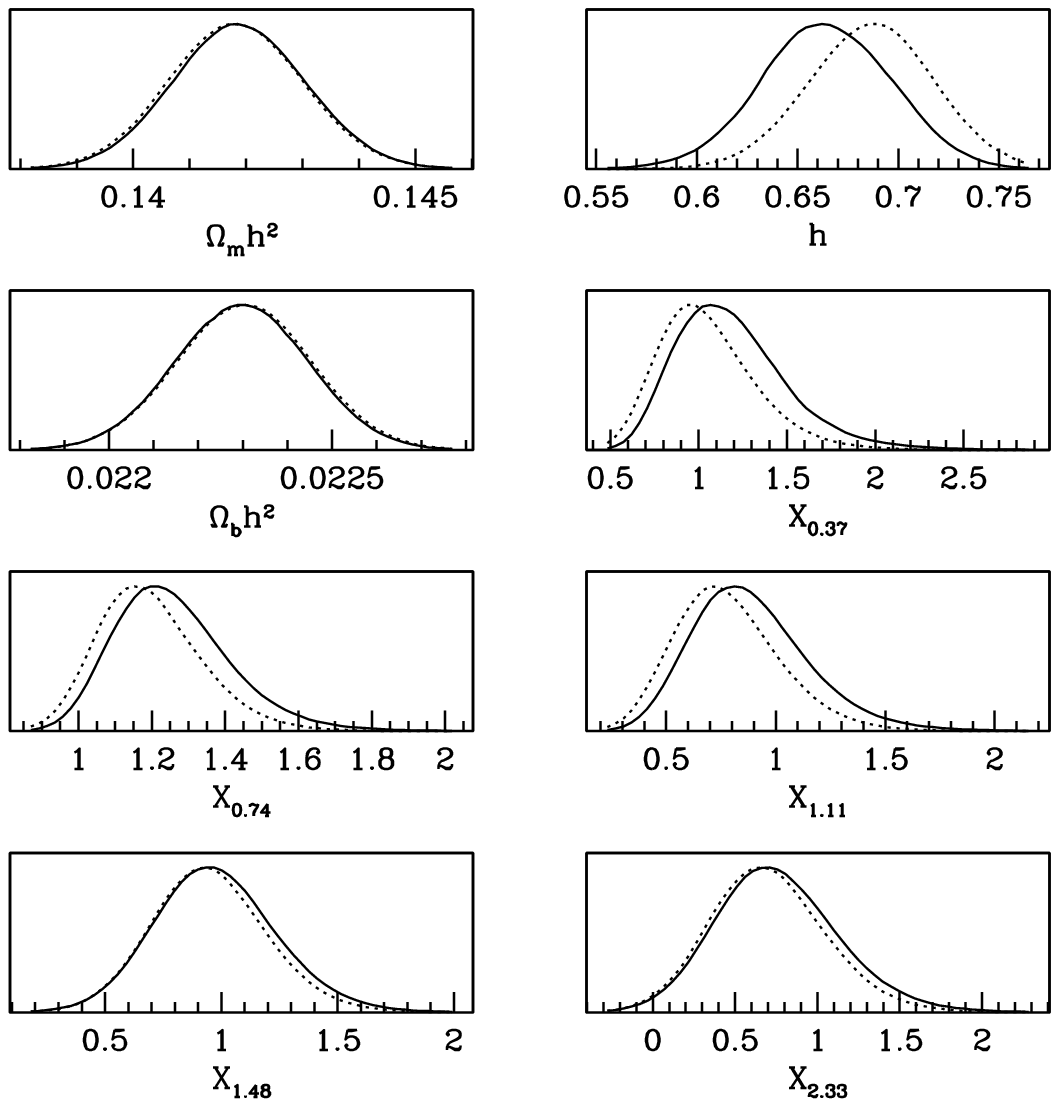}
\includegraphics[width=0.495\columnwidth,clip]{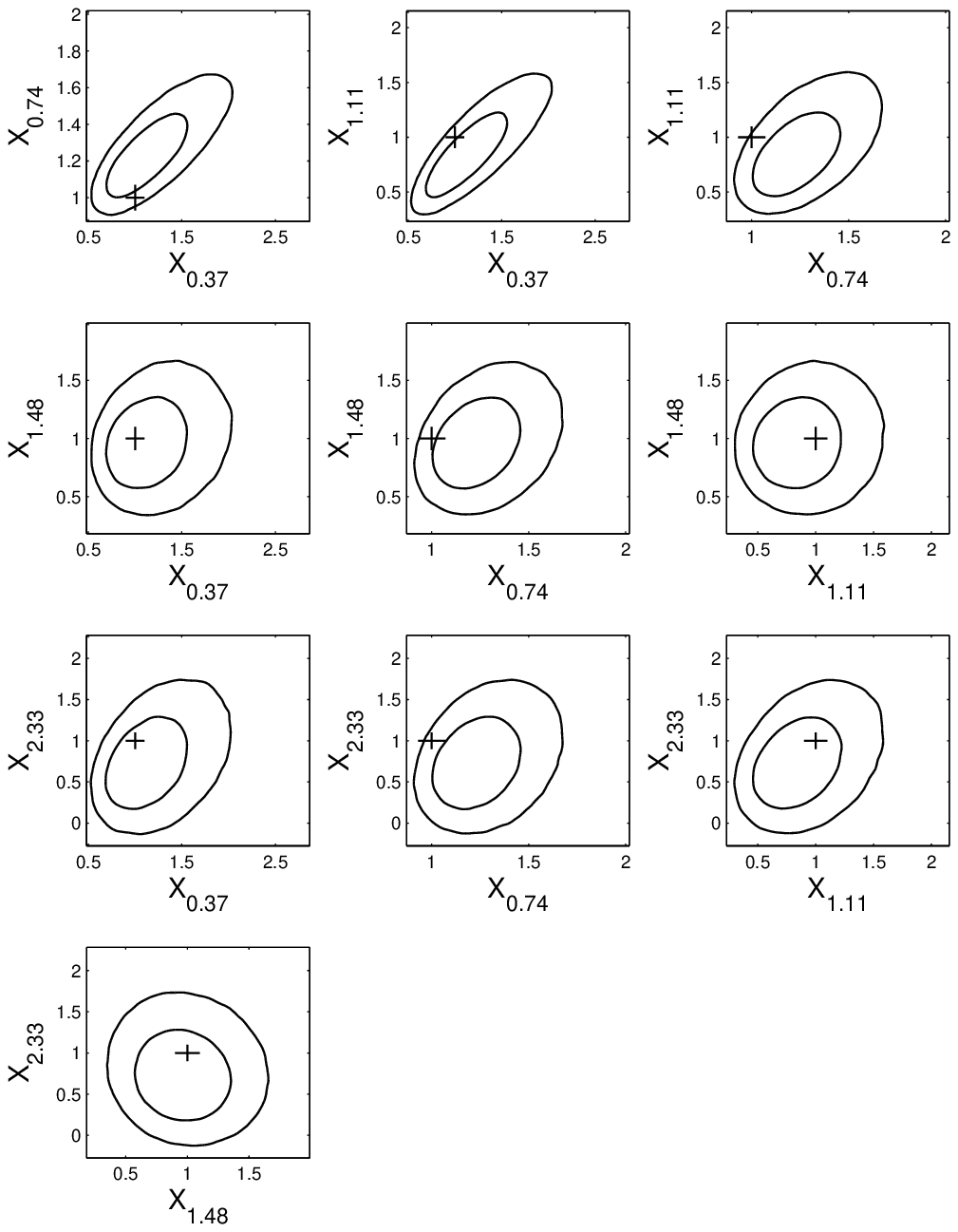}
\caption{Same as Fig.8, but for measuring $\rho_X(z)$ using DESI DR2 BAO measurements and Planck priors from \cite{WangDai2016} at $\{z_i\}=\{0.37, 0.74, 1.11, 1.48,2.33\}$.  Probability density distributions of  cosmological and dark energy parameters (left) and the corresponding 68.3\% and 95.4\% joint confidence level contours of the dark energy parameters (right), with 
the small cross representing the cosmological constant. } 
\label{fig:alt-dz0d37}
\end{figure}

The pdf of the estimated parameters and the joint confidence level contour plots of the dark energy parameters corresponding to Fig.\ref{fig:Xz-dz} are shown in 
in Fig.\ref{fig:alt-Xz4} for the $X(z)$ parametrization with four parameters with d$z=1/3$ (left panels, AIC/BIC preferred) and d$z=0.45$ (right panels), and in
Fig.\ref{fig:alt-dz0d37} for the alternative case of five $X(z)$ parameters with d$z=0.37$.
These are very similar to those for our fiducial case of d$z=1/3$ in Fig.2-5, indicating the robustness of our approach.

To summarize, here is the recipe for optimizing the $X(z)=\rho_X(z)/\rho_X(0)$ measurement:\\
\noindent
(1) Divide the redshift range with sufficient data into a set of evenly distributed redshift values, $\{z_i\}=(i-1)\mbox{d}z$ ($i=1, ..., N$), and $z_{N+1}$ is the maximum redshift from the data. \\
\noindent
(2) For each choice of $\{z_i\}$, carry out an MCMC analysis with $\{X(z_{i+1})\}$ ($i=1, 2, ..., N)$ as dark energy parameters (noting that $z_1=0$ and $X(z_1)=1$), and obtain $X(z)$ using cubic-spline from $\{X(z_i)\}$ ($i=1, 2, ..., N+1)$. \\
\noindent
(3) Find the set of $\{z_i\}$ with the smallest d$z$ that has $\{X(z_{i+1})\}$ ($i=1, 2, ..., N)$ well constrained by data.\\
\noindent
(4) Remove $X(z_j)$ if it contains no additional information, i.e., it can be inferred from the two adjacent measurement, $X(z_{j-1})$ and $X(z_{j+1})$.\\
\noindent
(5) Compute AIC and BIC for the analysis with $X(z)$ parameterized by $N$ parameters, and that of a smaller number of parameters (with the redundant $X(z)$ parameters removed, as described in step (4)), to confirm that the reduced set represents the optimal $X(z)$ measurement preferred by AIC and BIC.

We followed this procedure and studied the four different sets of $\{z_i\}$ for parametrizing $X(z)$ listed at the beginning of this section, and found little sensitivity to these choices in the resultant $X(z)$ measured.  Removing the parameter point at $z=4/3$ from the fiducial case in Fig.1 leads to the lowest AIC and BIC values (i.e., preferred by model selection, see Table 2) but with very similar $X(z)$ as previously obtained in Fig.1 (see Fig.\ref{fig:Xz-dz}).

\subsection{Constraints on \texorpdfstring{$(w_0,w_a)$}{TEXT}}
\label{sec:w0wa}

For comparison with the work by others, we have also derived constraints on $(w_0,w_a)$ by combining DESI DR2 BAO measurements with two different ways of adding compressed Planck CMB constraints:\\
\noindent
(1) $\{R,l_a,\Omega_bh^2\}$ as proposed by one of us in \cite{WangPia2007}, from \cite{WangDai2016}.\\
\noindent
(2) $\{\theta_*, \Omega_bh^2, \Omega_{bc} h^2\}$ from \cite{Lemos2023}, with $\Omega_{bc}=\Omega_b+\Omega_{cdm}$ as \cite{DESI-DR2}. 

Each set of CMB constraints are added as Gaussian priors, with their correlations properly accounted for (see Eq.[18] in \cite{WangDai2016}).
Table 3 lists our constraints on ($w_0,w_a$) from combining DESI DR2 BAO measurements with each of these two compressed Planck CMB data sets. In the fourth column of the table, $r$ is the correlation coefficient of $w_0$ and $w_a$. One can see that they are always highly correlated.
\begin{table}
\begin{footnotesize}
\begin{tabular}{|l|l|l|l|l|l|}\hline
Planck CMB data & $w_0$ & $w_a$ & $r(w_0,w_a)$\\
\hline
$\{R,l_a,\Omega_bh^2\}$ & & & \\
This paper &  $-0.46\pm  0.24$ & $-1.62\pm 0.71$ & $-0.982$\\
\hline
$\{\theta_*, \Omega_bh^2, \Omega_{bc} h^2\}$ & & & \\
(a) this paper & $-0.43\pm  0.23$ & $-1.71\pm 0.70$ & $-0.981$\\
%\hline
(b) DESI team \cite{DESI-DR2} & $-0.43\pm 0.22$ & $-1.72 \pm 0.64$
& (not provided)\\
\hline 
Full CMB data & & &\\
DESI team \cite{DESI-DR2} & $-0.42\pm 0.21$ & $-1.75 \pm 0.58$
& (not provided)\\
\hline
\end{tabular}
\caption{Constraints on ($w_0,w_a)$ from combining DESI DR2 BAO measurements with Planck CMB data. Note that $r$ is the correlation coefficient of $w_0$ and $w_a$.
One can see that $w_0$ and $w_a$ are always highly correlated.}
\label{tab:w0wa}
\end{footnotesize}
\end{table}

We find consistent constraints from the two different ways of compressing CMB data. We reexamined the same compressed CMB data set $\{\theta_*, \Omega_bh^2, \Omega_{bc} h^2\}$ used by the DESI team, and we find essentially the same marginalized mean values but slightly larger errors ($\sim $ 10\% larger). 
They find that using the full Planck CMB data set shrinks the errors by $\sim$ 10\% due to the inclusion of CMB lensing (see also \cite{Park2024})
and thus claim 3.1$\sigma$ deviation in ($w_0,w_a$) from $\Lambda$CDM. 
Given the nature of statistical analysis, it is difficult to attach a great deal of significance to the small differences in Table \ref{tab:w0wa}.

\section{Comparison With Previous Work}
\label{sec:compare}

Subsequent to the original DESI DR2 papers, many authors have reexamined the evidence for time-varying dark energy in the data, e.g., \cite{Shlivko2405,Shajib2502,Shlivko2504,Cortes2504,Choudhury2025,Dinda2025,Efstathiou2505,Park2025,Nesseris2025, Cheng2025} as well as many others (see also, e.g. \cite{Choudhury2024,Dinda2024,Payeur2024,Wolf2024}, for analysis using DESI DR1 data).
In this section, we compare our work with the follow-up approaches of the DESI Collaboration in their extended dark energy analysis paper \cite{DESI:2025fii}, and discuss the major differences.
Note that we have not included supernova (SN) data in our results and instead restricted to DESI + CMB (or priors on $\Omega_mh^2$), to avoid introducing complications arising from the differences among the different SN data sets. 

In \cite{DESI:2025fii}, the DESI collaboration extended their analysis to a variety of alternative dark energy models beyond the original linear model and claimed that these other approaches all validate the $w_0w_a$ results from their original study in \cite{DESI-DR2}. 
First they reproduced the same results as their first paper using four alternative $w_0 w_a$ parameterizations.  
Secondly they used an expansion in terms of Chebyshev polynomials; their Figure 5 shows results similar to ours; they find DESI + CMB consistent with $\Lambda$CDM at $\sim 2 \sigma$ (their $X(z)$ is consistent with $\Lambda$ at 2$\sigma$ at all redshifts; their $w_X(z)$ contours for DESI+CMB are obscured by their DESI+CMB+SN contours at $0.3<z<0.5$, but consistent with $\Lambda$ at other redshifts at 2$\sigma$), smaller than the $3.1 \sigma$ discrepancy they found using the $w_0w_a$ parametrization for the same data.  They focus their discussion on the fact that their Chebyshev results are consistent with their $w_0w_a$ analysis, which is true, but
in fact their results are also not a bad fit to $\Lambda$CDM, although they do not compute that fit.
Thirdly, they apply two non-parametric approaches: binning, and Gaussian process regression (GP). 

Gaussian process regression (GP) samples the parameter space randomly, assuming a mean function, with a kernel that controls the correlation length of the samples, and the typical deviations from the mean function. 
The results from GP in the second DESI paper \cite{DESI:2025fii} for the statistical significance for time-varying DE using DESI+CMB are not very different from ours.
As can be seen in their Figure 9 (top left panel), their error bars on $w(z)$ grow with redshift (sensibly becoming very large at high redshift where there are no data).
Again, their discussion focuses on the consistency of their GP method results with those from their $w_0w_a$ method where, in fact, the deviation from $\Lambda$CDM appears to be just over $2 \sigma$ in their GP reconstructed $w_X(z)$, consistent with our results on $w_X(z)$.
It is worth noting that the $\rho_X(z)$ associated with their GP reconstructed $w(z)$ (see their Fig.9, left panel) deviates by more than 2$\sigma$ from a constant at $z \ga 2.33$, where there is no data. This is a generic implication of $w_X(z)$ measurements: boundary conditions placed on $w_X(z)$ at high $z$ implies big assumptions about $\rho_X(z)$ unjustified by the data.

In the binned analysis of the data by \cite{DESI:2025fii}, they assume: 
\begin{equation}
    w(z) = w_0 + \sum_{i=1}^N \frac{(w_i - w_{i-1})}{2} \left[1 + {\rm tanh}\biggl(\frac{z-z_i}{s} \biggr) \right] 
    \label{eq:DESI_wz}
\end{equation}
where $w_i$ are the bin amplitude parameters, $N$ is the number of bins, and $s$ is the smoothing scale.
This approach is somewhat similar to ours (they took smoothed steps where we took cubic spline in between points $w_i$).
Here are the main differences between their results and ours:
(1) They only showed results for DESI + CMB + SN.
We have shown results from DESI + CMB, without including SN, which is far more robust at present.
(2) They binned the data (Eq.[\ref{eq:DESI_wz}] is effectively a smoothed step function based on $\{w_i\}$), while we interpolate $w_X(z)$ and $X(z)$ smoothly using cubic spline from $\{w_i\}$ and $\{X_i\}$, which joins adjacent points using cubic polynomials, thus the two are similar yet different: a step function is less physical than a smooth function without artificial steps.
(3) Our choice of $\{z_i\}$ is closer to where the data are (see Sec.3.2).
The DESI bins are uniform out to $z=2.1$, even though there is only one BAO measurement at $z>1.5$.
Our redshift values for the dark energy measurements are uniformly distributed only at $0< z \la 1.33$, the range where most of the DESI BAO measurements are, plus $z=2.33$ (the highest redshift data point).
However, the $\{z_i\}$ choices are not likely to play a major role in the conclusions.

Their Fig.7 shows binned $w_X(z)$ and $X(z)$ from DESI+CMB+Union3, which are inconsistent with each other at $0.5<z<1$: their binned $w_X(z)$ is consistent with $\Lambda$CDM well within 2$\sigma$, but their $X(z)$ deviates from $\Lambda$CDM at $\sim 3 \sigma$.
Further, there is another puzzling discrepancy between these two results from the DESI collaboration.
On the one hand, at $z<0.5$, there is only one distance measurement from DESI BGS, and that measurement (at $z=0.295$, see the top left panel of Fig.6 in \cite{DESI-DR2}) is consistent with $\Lambda$CDM. Yet, in the same redshift region $z< 0.5$, both of DESI's binned $w_X(z)$ and $\rho_X(z)$ deviate from $\Lambda$CDM at $>2\sigma$.

Note that our results do agree qualitatively with that of the DESI Collaboration, but at a significantly reduced statistical significance: we find less than 2$\sigma$ deviation, while they find 3$\sigma$ deviation (assuming a linear dark energy equation of state), from a cosmological constant.
This is important since 3$\sigma$ is commonly regarded as the threshold for ``evidence'' for a possible discovery, while 2$\sigma$ is still considered to be within statistical uncertainties.
It is important that Euclid and Roman will provide data points in the current data gap at $z>1.5$ (see Fig.1 \blue{and Fig.7}), which will help clarify whether dark energy does vary with time, independent of model assumptions.

\section{Summary and Discussion}
\label{sec:summary}

In this paper, we have measured the dark energy density $\rho_X(z)$ and equation of state $w_X(z)$ from data in a truly model-independent fashion.
Using DESI DR2 BAO distance measurements \cite{DESI-DR2} and Planck CMB distance priors \cite{WangDai2016}, we measured both as free functions of redshift $z$, parametrized by their values at $z=0, 1/3, 2/3, 1, 1.33, 2.33$ (we smoothly interpolated between these redshifts using cubic spline).
Our main result is that for a flat Universe (assumed in the DESI DR2 BAO measurement), we find $\rho_X(z)$ and $w_X(z)$ to be 
consistent with a cosmological constant, with only a deviation at $\sim 1\sigma$ for $\rho_X(z)$ and $\sim 2\sigma$ for $w_X(z)$ at $z=2/3$ (see Fig.1). 
Note that current data do not place meaningful constraints on dark energy without assuming a flat Universe. 

In addition, we find that measuring $\rho_X(z)$ is preferred to measuring $w_X(z)$ by model selection using AIC and BIC (see Table 1), and that $w_X(z)$ is significantly less constrained by data than $\rho_X(z)$ (see Fig.1), confirming our results in \cite{WangFreese2006}. 

We tested the robustness of our $\rho_X(z)$ measurement by adding Planck distance 2015 and 2018 priors that are more relaxed (allowing $m_\nu$ to vary) from \cite{Zhai2020}, and find that the results are very similar to our baseline choice of Planck 2015 distance priors from \cite{WangDai2016} (fixing $m_\nu=0.06\,$eV), with nearly identical deviation from a cosmological constant, $\sim 1\sigma$ at $z=2/3$. 
This indicates that \cite{WangDai2016} does give the correct and simplest way of adding Planck constraints.

We have also showed that varying the specific choice of redshifts of the $X(z)\equiv \rho_X(z)/\rho_X(0)$ measurement (away from the fiducial ones), as well as varying the number of $X(z)$ parameters, leads to very consistent results, see Fig.\ref{fig:Xz-dz}. The AIC and BIC values change less than two (thus all models are practically equivalent) but slightly favoring our fiducial case with the $X(z=4/3)$ parameter removed (see Table 2). All of these are in agreement with a cosmological constant except for the 1-2$\sigma$ deviation at $0.4 \la z \la 0.9$, the same redshift range where DESI DR2 BAO measurements deviate from a cosmological constant at about the same statistical significance (see Fig.6 of \cite{DESI-DR2}). This indicates that our method is effective in accurately translating the BAO measurements into dark energy density measurements.
We summarize how to obtain the optimized $X(z)$ measurement at the end of Sec.3.2.

The DESI Collaboration, on the other hand, finds that DESI DR2 data combined with Planck CMB data (the full data instead of compressed statistics) leads to 3.1$\sigma$ deviation from a cosmological constant (reduced to 2.4$\sigma$ for using compressed CMB data without CMB lensing) \cite{DESI-DR2}, assuming $w_X(z)=w_0+w_a(1-a)$. 
Our results differ primarily due to our model-independent approach of measuring $\rho_X(z)$ as a free function, as opposed to the $(w_0, w_a)$ parametrization used by the DESI collaboration.
This indicates that assuming a linear $w_X(z)$ is misleading, as it cannot capture the dark energy properties from data in a model-independent manner.  

It is worth noting that the best-fit $w_X(z)=w_0+w_a(1-a)$ models from DESI DR2 from \cite{DESI-DR2} all fall within the 1$\sigma$ contours of our model-independent measurements (see Fig.1 and Fig.7). Thus even though the DESI DR2 best-fit $w_X(z)=w_0+w_a(1-a)$ models deviate from a cosmological constant by more than 3$\sigma$ \cite{DESI-DR2}, they are consistent with dark energy density measured as a free function that deviates from a cosmological constant by only 1$\sigma$.
This means that the results from fitting $w_X(z)=w_0+w_a(1-a)$ must be interpreted with care. 

Our analysis has not included ACT CMB data, to supplement Planck out to smaller scales.
It is likely that adding ACT would lead to even less significance of tension vs. $\Lambda$CDM, for comparison see 
\cite{Calabrese2503,Garcia-Quintero2504} (still a subject of controversy).

We have focused on the DESI DR2 plus Planck CMB data only in this paper, in order to make the simplest comparison with the DESI Collaboration results in a straightforward and transparent manner, especially since they find a 3.1$\sigma$ deviation from a cosmological constant with only Planck CMB data added.
Adding supernova data is more complicated, since the different supernova data sets give somewhat different results, which is worthy of investigation in its own right, which we will leave to a future study.

We have found that the assumption about $w_X(z)$ at high $z$ (beyond the last DESI DR2 data point) has a significant impact on the resultant constraints on $w_X(z)$, in particular, how much it deviates from a cosmological constant.
Note that the most natural parametrization of $w_X(z)$ all trend to a constant $w$ at high $z$, which implies a dark energy density $\rho_X(z) \propto (1+z)^{3(1+w)} $.
This seems to be a fundamental limitation of parametrizing our ignorance of dark energy using its equation of state --- assuming $\rho_X(z) \propto (1+z)^{3(1+w)} $ at high $z$ is a big assumption about dark energy without any observational evidence. Removing this assumption requires the introduction of additional parameters, which introduce more assumptions about dark energy at high $z$.

In our quest to discover the physical nature of dark energy, the most urgent goal at present is to determine definitively whether dark energy density varies with time. Thus it is of critical importance to measure dark energy density from data in a model-independent manner. Dark energy equation of state is less constrained by data since it only enters theoretical prediction via the dark energy density $\rho_X(z)$, which is given by integrating over a function of $w_X(z)$ (see Eq.[\ref{eq:wX}]). 

Given the current data gap at $z>1.5$ (with a single BAO measurement from DESI Ly$\alpha$ at $z=2.33$), future data from Euclid at $0.9 \leq z \leq 1.8$ \cite{Euclid-overview,Wang2010} and Roman Space Telescope at $ 1 \leq z \la 3$ \cite{Spergel2015,Wang2022} will
significantly advance our understanding of dark energy (see Fig.1, left panel, and Fig.7).

\section*{Acknowledgements}

We thank Scott Dodelson, Will Percival, Hee-Jong Seo, and Gab Montefalcone for helpful discussions.
YW gratefully acknowledges funding from NASA Grant \#80NSSC24M0021, ``Project Infrastructure for the Roman Galaxy Redshift Survey''.
KF is grateful for support due to the position she holds as the
Jeff \& Gail Kodosky Endowed Chair in Physics.  She is also grateful for support from the Swedish Research Council (Contract No. 638-2013-8993).


\begin{thebibliography}{}

\bibitem{Riess98}
A. Riess, et al., 1998, AJ, 116, 1009
\bibitem{Perl99}
S. Perlmutter, et al., 1999, ApJ, 517, 565

\bibitem{Wang-DE-book}
Y. Wang, Dark Energy (Wiley-VCH, Weinheim, Germany,
2010).

\bibitem{Blake2003}
C. Blake; G. Glazebrook, 2003, ApJ, 594, 665
\bibitem{Seo2003}
H. Seo; D. Eisenstein, 2003, ApJ, 598, 720

\bibitem{Kaiser1987}
N. Kaiser, 1987, MNRAS 227, 1
\bibitem{Guzzo2008}
L. Guzzo, et al. 2008, Nature 451, 541
\bibitem{Wang2008a}
Y. Wang, 2008, JCAP, 5, 021
% arXiv:0710.3885
%Differentiating dark energy and modified gravity with galaxy redshift surveys

\bibitem{DESI-DR2}
DESI Collaboration, M. Abdul-Karim, et al., 2025a, arXiv:2503.14738
%"DESI DR2 Results II: Measurements of Baryon Acoustic Oscillations and Cosmological Constraints"
%arXiv:2503.14738

%\cite{DESI:2025fii}
\bibitem{DESI:2025fii}
DESI Collaboration, K. Lodha, \textit{et al.}, 2025b,
%``Extended Dark Energy analysis using DESI DR2 BAO measurements,''
arXiv:2503.14743 

\bibitem{WangFreese2006}
Y. Wang, and K. Freese, Phys. Lett. B 632, 449 (2006).
%astro-ph/0402208 
%Probing Dark Energy Using Its Density Instead of Its Equation of State

\bibitem{ParticleDataGrp2022}
R. L. Workman, V. D. Burkert, V. Crede, E. Klempt,
U. Thoma, L. Tiator, K. Agashe, G. Aielli, B. C. Allanach,
C. Amsler, M. Antonelli, E. C. Aschenauer,
D. M. Asner, H. Baer, S. Banerjee, and others, Progress
of Theoretical and Experimental Physics 2022, 083C01
(2022).

\bibitem{Neutrino-review2012}
J. Lesgourgues and S. Pastor, Adv. High Energy Phys.
2012, 608515 (2012), arXiv:1212.6154 [hep-ph].

\bibitem{WangTegmark2004}
Y. Wang, and M. Tegmark, Phys. Rev. Lett. 92, 241302
(2004).
% astro-ph/0403292
%New dark energy constraints from supernovae, microwave background and galaxy clustering

\bibitem{WangPia2007}
Y. Wang, and P. Mukherjee, Phys. Rev. D 76, 103533 (2007).
%astro-ph/0703780
%Observational Constraints on Dark Energy and Cosmic Curvature

\bibitem{WangDai2016}
Y. Wang, and M. Dai, PRD, 94, 083521 (2016)
%Exploring uncertainties in dark energy constraints using current
%observational data with Planck 2015 distance priors
%arXiv:1509.02198

\bibitem{Zhai2020}
Z. Zhai, C.-G. Park, Y. Wang, B. Ratra, JCAP, 07, 009 (2020),
%CMB distance priors revisited: effects of dark energy dynamics, spatial curvature, primordial power spectrum, and neutrino parameters
%arXiv:1912.04921 

\bibitem{Aizpuru2021}
A. Aizpuru, R. Arjona, S. Nesseris, Phys. Rev. D 104, 043521 (2021)
%Machine Learning improved fits of the sound horizon at the baryon drag epoch
%2106.00428

\bibitem{Hu96}
W. Hu, and N. Sugiyama, Astrophys. J. 471, 542 (1996).

\bibitem{Lewis2002}
A. Lewis, and S. Bridle, Phys. Rev. D 66, 103511 

\bibitem{Akaike1974}
H. Akaike, IEEE Trans. Autom. Control 19, 716 (1974). doi:10.
1109/TAC.1974.1100705

\bibitem{Schwarz1978}
G. Schwarz, Ann. Stat. 6, 461 (1978). doi:10.1214/aos/
1176344136

\bibitem{Lemos2023}
P. Lemos and A. Lewis, Phys. Rev. D 107, 103505 (2023)

\bibitem{Freese1987}
K. Freese, F.C. Adams, J.A. Frieman, E. Mottola, 
Nuclear Physics B, 287, 797 (1987)
%Cosmology with decaying vacuum energy

\bibitem{Dodelson2000}
S. Dodelson; M. Kaplinghat; E. Stewart, 2000, PRL, 85, 5276
% arXiv:astro-ph/0002360
%"Solving the Coincidence Problem: Tracking Oscillating Energy"

\bibitem{Park2024}
C.-G. Park, J. C. Perez, B. Ratra, arXiv:2410.13627 

\bibitem{Shlivko2405}
D. Shlivko, P.J. Steinhardt, arXiv:2405.03933, 
%Assessing observational constraints on dark energy

\bibitem{Shajib2502}
A.J. Shajib, and J.A. Frieman, arXiv:2502.06929, 
%Evolving dark energy models: Current and forecast constraints

\bibitem{Shlivko2504}
D. Shlivko.; P.J. Steinhardt; C.L. Steinhardt, arXiv:2504.02028
%Optimal parameterizations for observational constraints on thawing dark energy

\bibitem{Cortes2504}
M. Cortes, and A.R. Liddle, arXiv:2504.15336
%On DESI's DR2 exclusion of ΛCDM

\bibitem{Choudhury2025}
S.R. Choudhury, 2025 ApJL 986 L31
%Cosmology in Extended Parameter Space with DESI Data Release 2 Baryon Acoustic Oscillations: A 2σ+ Detection of Nonzero Neutrino Masses with an Update on Dynamical Dark Energy and Lensing Anomaly, 	arXiv:2504.15340

\bibitem{Dinda2025}
B.R. Dinda, R. Maartens, S. Saito, C. Clarkson, arXiv:2504.09681 
% Improved null tests of ΛCDM and FLRW in light of DESI DR2

\bibitem{Efstathiou2505}
G. Efstathiou, arXiv:2505.02658
%Baryon Acoustic Oscillations from a Different Angle

\bibitem{Park2025}
C.-G. Park, B. Ratra, arXiv:2501.03480 

\bibitem{Nesseris2025}
S. Nesseris, Y. Akrami, G.D. Starkman, arXiv:2503.22529
% To CPL, or not to CPL? What we have not learned about the dark energy equation of state

\bibitem{Cheng2025}
H. Cheng, et al., arXiv:2505.02932
%Hanyu Cheng, Eleonora Di Valentino, Luis A. Escamilla, Anjan A. Sen, Luca Visinelli, Pressure Parametrization of Dark Energy: First and Second-Order Constraints with Latest Cosmological Data

\bibitem{Choudhury2024}
S.R. Choudhury, \& T. Okumura, 2024 ApJL 976 L11
% Updated Cosmological Constraints in Extended Parameter Space with Planck PR4, DESI Baryon Acoustic Oscillations, and Supernovae: Dynamical Dark Energy, Neutrino Masses, Lensing Anomaly, and the Hubble Tension, 	arXiv:2409.13022

\bibitem{Dinda2024}
B.R. Dinda, \& R. Maartens, JCAP01(2025)120
% Model-agnostic assessment of dark energy after DESI DR1 BAO, 	arXiv:2407.17252

\bibitem{Payeur2024}
G. Payeur, E. McDonough, \& R. Brandenberger, 2025, PRD, 111, 3541
%Do Observations Prefer Thawing Quintessence?, 	arXiv:2411.13637

\bibitem{Wolf2024}
W.J. Wolf, C. García-García, D.J. Bartlett, P.G. Ferreira, Phys. Rev. D 110 (2024) 083528
% Scant evidence for thawing quintessence, 	arXiv:2408.17318 

\bibitem{Calabrese2503}
E. Calabrese, et al., arXiv:2503.14454, 
%The Atacama Cosmology Telescope: DR6 Constraints on Extended Cosmological Models

\bibitem{Garcia-Quintero2504}
C. Garcia-Quintero, et al., arXiv:2504.18464
%Cosmological implications of DESI DR2 BAO measurements in light of the latest ACT DR6 CMB data


\bibitem{Euclid-overview}
Euclid Collaboration: Y. Mellier, et al., 2025, A\&A, 697A, 1
%Euclid. I. Overview of the Euclid mission
% arXiv:2405.13491

\bibitem{Wang2010}
Y. Wang, et al., MNRAS, 409, 737 (2010)
%"Designing a space-based galaxy redshift survey to probe dark energy"
% arXiv:1006.3517

\bibitem{Spergel2015}
D.N. Spergel, et al. 2015, WFIRST SDT final report, arXiv:1503.03757

\bibitem{Wang2022}
Y. Wang, et al., ApJ, 928, 1 (2022)
%The High Latitude Spectroscopic Survey on the Nancy Grace Roman Space Telescope, 
% arXiv:2110.01829

\end{thebibliography}
\end{document}